\documentclass[conference]{IEEEtran}
\IEEEoverridecommandlockouts
\usepackage[utf8]{inputenc} 
\usepackage[T1]{fontenc}    
\usepackage[utf8]{vietnam}
\usepackage[english]{babel}
\usepackage{tipa}
\usepackage{ragged2e}

\usepackage{cite}
\usepackage{amsmath,amssymb,amsfonts}
\usepackage{algorithmic}
\usepackage{graphicx}
\usepackage{textcomp}
\usepackage{xcolor}

\usepackage{tabularx}
\usepackage{tkz-tab}

\usepackage{pgfplots}
\usepackage{tikz,pgfplots,pgfplotstable}

\usepackage{framed,multicol}
\usepackage[framemethod=tikz]{mdframed}

\usepgfplotslibrary{colorbrewer}
\pgfplotsset{compat = 1.15, cycle list/Set1-8} 
\usetikzlibrary{pgfplots.statistics, pgfplots.colorbrewer} 
\usepackage{pgfplotstable}
\usepackage{filecontents}

\definecolor{markercolor}{RGB}{124.9, 255, 160.65}
\pgfplotsset{compat=1.3}
\usetikzlibrary{positioning, fit, calc}  
\tikzset{block/.style={draw, thick, text width=2cm ,minimum height=1.3cm, align=center},   
	line/.style={-latex}     
} 

\tikzset{blocktext/.style={draw, thick, text width=5.2cm ,minimum height=1.3cm, align=center},   
	line/.style={-latex}     
}

\def\BibTeX{{\rm B\kern-.05em{\sc i\kern-.025em b}\kern-.08em
    T\kern-.1667em\lower.7ex\hbox{E}\kern-.125emX}}

\makeatletter
\def\endthebibliography{%
	\def\@noitemerr{\@latex@warning{Empty `thebibliography' environment}}%
	\endlist
}
\makeatother

\begin{document}

\title{Evaluation of ChatGPT and Microsoft Bing AI Chat Performances on Physics Exams of Vietnamese National High School Graduation Examination \\
}

\author{\IEEEauthorblockN{Xuan-Quy Dao}
\IEEEauthorblockA{\textit{School of Engineering} \\
\textit{Eastern International University}\\
Binh Duong, Vietnam \\
quy.dao@eiu.edu.vn}
\and
\IEEEauthorblockN{Ngoc-Bich Le}
\IEEEauthorblockA{\textit{School of Biomedical Engineering } \\
\textit{International University Vietnam, National University HCM City}\\
HMC City, Vietnam \\
lnbich@hcmiu.edu.vn }
\and
\IEEEauthorblockN{Xuan-Dung Phan}
\IEEEauthorblockA{\textit{School of Engineering} \\
\textit{Eastern International University}\\
Binh Duong, Vietnam \\
dung.phan@eiu.edu.vn}
\and
\IEEEauthorblockN{Bac-Bien Ngo}
\IEEEauthorblockA{\textit{School of Engineering} \\
\textit{Eastern International University}\\
Binh Duong, Vietnam \\
ngobacbienspk@gmail.com}
\and
\IEEEauthorblockN{The-Duy Vo}
\IEEEauthorblockA{\textit{School of Engineering} \\
\textit{Eastern International University}\\
Binh Duong, Vietnam \\
duy.vo@eiu.edu.vn}
}

\maketitle

\begin{abstract}

The promise and difficulties of language model-based approaches for physics teaching were assessed in this study. This study evaluates how well ChatGPT and BingChat, two state-of-the-art (SOTA) large language models (LLMs), perform when answering high school physics questions on Vietnamese exams from 2019 to 2023. When we compared the results of the LLMs with the scores of Vietnamese students, we discovered that ChatGPT and BingChat both perform worse than Vietnamese students, proving that LLMs are not yet capable of fully replacing human intellect in the field of physics teaching. The outcomes also showed that neither LLM is capable of responding to questions at the high application levels. In terms of accuracy, BingChat typically surpassed ChatGPT, although ChatGPT showed more stability. Our research suggests that LLMs can help students and teachers during learning and teaching activities, particularly by offering immediate feedback and individualized learning experiences

\end{abstract}


\section{Introduction}

Artificial intelligence (AI) integration into educational settings has grown in popularity in recent years with the goal of strengthening student learning and teaching methods. Automating repetitive tasks, offering real-time feedback and assessment, and personalizing learning experiences are all capabilities of AI-powered educational systems. In a study on the effects of AI on education, Chen et al.~\cite{chen2020artificial} concentrated on the use of AI in administration, instruction, and learning to allow instructors to perform administrative functions more effectively and customize content based on students' needs, thereby improving the overall quality of learning. Furthermore, Dao et al.~\cite{Dao2021} discussed the use of AI in education to reduce workload and enhance learner engagement in online learning. Their approach involves using text-to-speech and speech-driven-face technology to automatically create a video lecture with the instructor's voice and face, eliminating the need for recording video and allowing for easy modification. In addition, Nguyen et al.~\cite{Nguyen2021} proposed an online learning platform that incorporates a Vietnamese virtual assistant to assist instructors in presenting lessons and assessing learners. The platform delivers lesson content through slides combined with a synthesized voice and the instructor's face, which enables easy editing without the need for video recording.

LLMs is a technology to create chatbots applied to education. LLMs models have demonstrated great potential in several applications, including language translation, content creation, and education. In 2018, Google introduced BERT~\cite{devlin2018bert} a pre-trained model that utilizes the Transformer architecture and has achieved impressive results in various natural language processing (NLP) tasks by being trained on an extensive corpus of text. RoBERTa~\cite{liu2019roberta}, introduced by Facebook in 2019, is an extension of BERT that uses a similar architecture but is trained on a larger corpus of text with longer sequences and more iterations. Another large language model, T5~\cite{raffel2020exploring},  was introduced by Google researchers in 2019. T5 employs a unified text-to-text approach, converting all tasks to text-to-text format and training them in a single model. OpenAI's GPT-3~\cite{brown2020language}, released in 2020, can perform various NLP tasks with minimal examples, earning recognition for its impressive performance.

To train, test and evaluate LLMs models, we need datasets. Several physics datasets evaluate the physics ability of LLMs. AI2 Reasoning Challenge–Physics dataset~\cite{clark2018think}, a multiple-choice question-answering dataset including questions from grade 3 to grade 9 science exams, a supporting knowledge base of 14.3M unstructured text passages. PhysNet Dataset~\cite{unke2019physnet} was designed for predicting energies, forces, and dipole moments of chemical systems using deep neural networks. ScienceQA~\cite{lu2022learn} has richer domain diversity than previous datasets, covering natural science, language science, and social science. ScienceQA features 26 topics, 127 categories, and 379 skills, categorizing questions by topic, category, and skill. These datasets challenge LLMs to demonstrate their physics ability.

LLMs' potential and difficulties in education are becoming more and more clear as they advance. However, it is essential to carry out thorough assessments of their capabilities, particularly in the area of high school physics, in order to successfully integrate these models into education, particularly in Vietnam where Vietnamese is the primary language. Despite this, there hasn't been any research on the subject, and there aren't many datasets that can be used to evaluate LLMs in high school physics. To bridge this gap, we have created the VNHSGE dataset~\cite{xuan2023vnhsge}, which contains data from the Vietnamese National High School Graduation Examination covering nine subjects, including physics. The dataset contains 19K multiple-choice questions and 300 essays on literature, featuring both text and images, and is available in JSON and Word formats.

In this paper, we focus on evaluating LLMs capacities on physic exams. The current study makes a number of contributions, including (1) a thorough evaluation of the performance of two SOTA-LLMs, ChatGPT and BingChat, in the context of high school physics education in Vietnam; (2) a comparison analysis of ChatGPT and BingChat's performance compared to Vietnamese students; and (3) an extensive investigation of the benefits and drawbacks of utilizing LLMs in the field of physics education in Vietnam.

\section{Related works}

\subsection{Large Language Models}

Based on GPT-3.5, OpenAI trained the very sophisticated big language model known as ChatGPT using a sizable corpus of text data. It has the potential to be employed in many educational applications and can produce human-like replies to input in natural language. It might be used, for instance, to create intelligent tutoring programs that offer individualized feedback, automate grading, and offer assessment. Additionally, ChatGPT could develop interesting educational materials on a variety of subjects to supplement already-existing teaching resources or even develop entirely new courses. On the other hand, Microsoft's BingChat  is a chatbot function of the Bing search engine. It has the potential to be a great educational tool because it can provide correct facts and content like stories and poems. By locating pertinent material and instructional resources, as well as producing content that can be incorporated into lesson plans, BingChat has the capacity to assist students with their homework and teachers with lesson preparation. Because of this, BingChat is a useful tool for both educators and students.

\subsection{Evaluation of LLMs on Physics}

LLMs have great natural language understanding skills because of their extensive training with large amounts of data. They are therefore the best candidates for meeting academic and professional standards. 
Lehnert et al.~\cite{lehnert2023ai} explored the abilities of ChatGPT to explain and explore theoretical physics concepts. ChatGPT is good at explaining ideas in different ways, but ChatGPT still has some limitations in physics because it can also confidently give out false information and statements. However, ChatGPT can help advance theoretical physics. 

Similarly, Kortemeyer et al.~\cite{kortemeyer2023could} studied ChatGPT's ability to pass an introductory physics course. ChatGPT almost passed despite numerous suppositions and errors. This highlights the importance of exercising caution and calls into question the reliability of AI-generated answers. 

In another study, West et al.~\cite{west2023ai} analyzed the performance of ChatGPT-3.5 and ChatGPT-4 in first-semester university physics using a modified version of the FCI. They discovered that, despite having inconsistent performance, ChatGPT-3.5 can match or outperform the median performance of a university student. The performance of ChatGPT-4 on topics pertaining to fundamental mechanics is comparable to that of a professional physicist.

The study by Kuchemann et al.~\cite{kuchemann2023physics} explored the potential of using ChatGPT 3.5 for physics task development by prospective teachers. In a randomized controlled trial with 26 physics teacher students, text-based physics activities for high school students were created using ChatGPT 3.5 in one group and a textbook in the other. Despite not finding a difference in task correctness between the two groups, the study did find that the textbook group had more clarity and better contextualization. The study emphasizes the benefits and drawbacks of utilizing extensive language models in instruction.

Yeadon et al.~\cite{yeadon2023death} presented evidence of AI-generated short-form physics essays achieving first-class grades in an accredited university physics module. The study discovered that 50 AI responses produced submissions with an average grade of 71$\%$ on an essay writing evaluation, in strong accord with the existing module average, using NLP AI such as ChatGPT and davinci-003. The study contends that the efficacy of short-form essays as an assessment tool in physics courses is seriously threatened by the most recent AI language models. The AI-generated essays had a low plagiarism score, proving they were original, according to plagiarism detection software. 

According to the GPT-4 Report by OpenAI~\cite{OpenAI_gpt_4_report}, ChatGPT-3.5 has an accuracy range of 33 to 66$\%$ on the AP Physics dataset.

This indicates that although ChatGPT has the potential to transform education, further efforts are necessary to enhance its precision in specific fields like Physics. 

\section{Dataset}

We use VNHSGE dataset~\cite{xuan2023vnhsge}, were taken from real exams and illustrative examples that were given from 2019 to 2023. It was compiled from information gathered from teachers, high schools, and the Vietnamese Ministry of Education and Training, among other places.

\subsection{Physics Testing of Vietnamese High School Graduation Examinations}
 
In Vietnam, the natural sciences combination includes the physics graduation exam, which is a significant component of the high school graduating test. For this test, students have 50 minutes to respond to 40 questions.

\subsection{Question Levels}

VNHSGE dataset contains a range of questions that assess various levels of complexity, from fundamental knowledge to challenging tasks that demand information processing and synthesis. In order to give a thorough assessment of students' proficiency and expertise, the questions were then divided into four difficulty levels: "knowledge (easy)","comprehension (intermediate)","application (difficult)", and "high application (very difficult). This classification strategy offers a thorough assessment of the LLMs' advantages and disadvantages in dealing with various problem types in physics.

\subsection{Question Topics}

A dataset for physics was used in this study that included 2000 multiple-choice questions divided into 50 sets of test questions. The questions include a wide range of physics topics, such as atomic nucleus, mechanical oscillations, mechanical waves, alternating current, electromagnetic oscillations and waves, light waves, quantum of light, electric charge and field, direct current, electromagnetic induction, and light refraction. These inquiries serve as a thorough evaluation of fundamental physics concepts and a test of students' comprehension.

\subsection{Score spectrum of Vietnamese students in 2019-2022}

A score distribution is a visual representation of how applicants performed in a specific subject. Scores are often shown on one axis of the chart, along with the number of applicants who received that score on the other axis. 

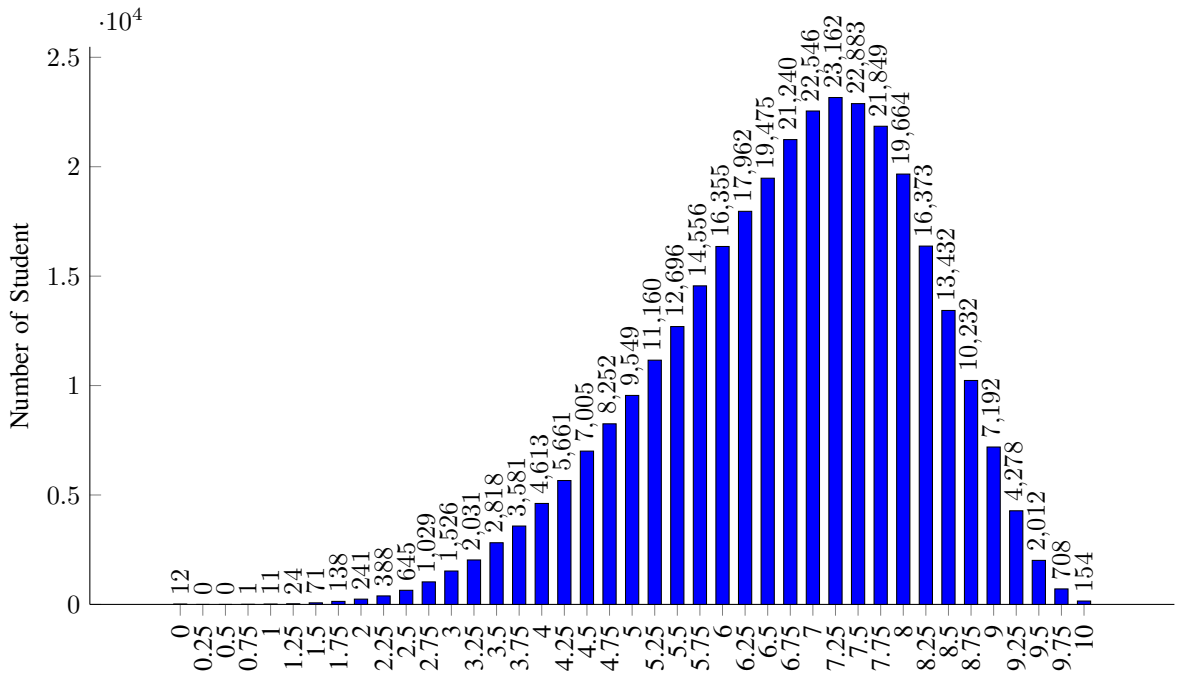
\begin{figure*}[ht!]
	\begin{center}
		\begin{tikzpicture}
			\begin{axis}[
				ylabel={Number of Student},
				symbolic x coords={
					0,
					0.25,
					0.5,
					0.75,
					1,
					1.25,
					1.5,
					1.75,
					2,
					2.25,
					2.5,
					2.75,
					3,
					3.25,
					3.5,
					3.75,
					4,
					4.25,
					4.5,
					4.75,
					5,
					5.25,
					5.5,
					5.75,
					6,
					6.25,
					6.5,
					6.75,
					7,
					7.25,
					7.5,
					7.75,
					8,
					8.25,
					8.5,
					8.75,
					9,
					9.25,
					9.5,
					9.75,
					10,	
				},
				xtick=data,
				x tick label style={rotate=90,anchor=east},
				ybar,
				bar width=5pt,
				ymin=0,
				nodes near coords,   
				every node near coord/.append style={rotate=90, anchor=west},
				width=\textwidth, 
				height=9cm, 
				width=16cm,
				axis x line*=bottom, axis y line*=left
				]
				\addplot [fill=blue] coordinates {
					(0,12)
					(0.25,0)
					(0.5,0)
					(0.75,1)
					(1,11)
					(1.25,24)
					(1.5,71)
					(1.75,138)
					(2,241)
					(2.25,388)
					(2.5,645)
					(2.75,1029)
					(3,1526)
					(3.25,2031)
					(3.5,2818)
					(3.75,3581)
					(4,4613)
					(4.25,5661)
					(4.5,7005)
					(4.75,8252)
					(5,9549)
					(5.25,11160)
					(5.5,12696)
					(5.75,14556)
					(6,16355)
					(6.25,17962)
					(6.5,19475)
					(6.75,21240)
					(7,22546)
					(7.25,23162)
					(7.5,22883)
					(7.75,21849)
					(8,19664)
					(8.25,16373)
					(8.5,13432)
					(8.75,10232)
					(9,7192)
					(9.25,4278)
					(9.5,2012)
					(9.75,708)
					(10,154)
					
				};	
				
			\end{axis}
		\end{tikzpicture}
	\end{center}
	\caption{Physics score spectrum of Vietnamese students in 2022.}
	\label{fig:phy_student_2022}
\end{figure*}

The analysis of the 2022 national high school graduation exam results in Physics, as shown in Fig.~\ref{fig:phy_student_2022}, revealed that 325,525 candidates took the Physics exam, with an average score of 6.72 points and the most attained score was 7.25 points. The score distribution, which is published annually by the Vietnamese Ministry of Education in chart form for each subject, is used to assess candidates' proficiency and ability, as well as to evaluate them based on predetermined criteria. The distribution is also used to assess and classify test papers according to difficulty level, allowing for the evaluation of candidates' quality. We collected score distributions from 2019 to 2022 to compare the performance of LLMs with that of Vietnamese students, providing insight into the capabilities of LLMs.

\subsection{VNHSGE dataset}
 
WORD format: The VNHSGE dataset is designed to be compatible with language models such as BERT and GPT, which require formulas, equations, and figures to be converted into text format. The dataset includes a WORD file in text format that can be easily evaluated by non-programmers. However, symbols, tables, and images are also converted into text format. The VNHSGE dataset is suitable for full language models like ChatGPT and BingChat.

JSON format: The JSON format is a fantastic choice for LLMs input data since it effectively handles both the syntax and substance of text data. Due to its adaptability and expansion, this format can store a variety of text data, including equations, formulas, tables, and images. The JSON format is a great fit for the VNHSGE dataset, making it compatible with a wide range of LLMs and offering a base for the creation of more reliable language models.

Samples: We will now present a set of questions that was translated from Vietnamese into English using ChatGPT and BingChat. Nonetheless, it is important to acknowledge that in certain instances, both models, particularly BingChat, may respond to Vietnamese questions in English. 

\subsubsection{Knowledge level question}

The first kind of question is at the knowledge level, and the solution can be determined without using any reasoning.

\begin{mdframed}[linewidth=1pt,linecolor=red]
\textcolor{red}{\textbf{Câu hỏi}}: Đặt điện áp xoay chiều có giá trị hiệu dụng U vào hai đầu một đoạn mạch chỉ có cuộn cảm thuần thì cảm kháng của đoạn mạch là $Z_L$. Cường độ dòng điện hiệu dụng I trong đoạn mạch được tính bằng công thức nào sau đây?

\textcolor{blue}{\textbf{Question}}: What is the formula for calculating the effective current I in a circuit consisting of a pure inductor with inductance $Z_L$ when an AC voltage with an effective value U is applied across the two ends of the circuit? 

A. $I=2UZ_L$	

B. $I=\frac{2U}{Z_L}$

C. $I=\frac{U}{Z_L}$	

D. $I=UZ_L$

\end{mdframed}

\subsubsection{Comprehension level question}

The following question requires a modest amount of inference to answer because it is at the comprehension level.

\begin{mdframed}[linewidth=1pt,linecolor=red]

\textcolor{red}{\textbf{Câu hỏi}}: Hai dao động điều hòa cùng tần số có pha ban đầu là $\varphi_1$ và $\varphi_2$. Hai dao động này cùng pha khi
	
\textcolor{blue}{\textbf{Question}}: Two harmonic oscillations with the same frequency and initial phases of $\varphi_1$ and $\varphi_2$. These oscillations are in phase when

A. $\varphi_2-\varphi_1=\left(2n+1\right)\pi$  with $n=\pm0,\pm1,\pm2,\ldots.$

B. $\varphi_2-\varphi_1=2n\pi$  with $n=\pm0,\pm1,\pm2,\ldots.$

C. $\varphi_2-\varphi_1=\left(2n+\frac{1}{5}\right)\pi$  with $n=\pm0,\pm1,\pm2,\ldots.$

D. $\varphi_2-\varphi_1=\left(2n+\frac{1}{3}\right)\pi$  with $n=\pm0,\pm1,\pm2,\ldots.$

\end{mdframed}

\subsubsection{Application level question}

The answer to the following question, which is at the application level, involves inference.

\begin{mdframed}[linewidth=1pt,linecolor=red]
	
	\textcolor{red}{\textbf{Câu hỏi}}: Trong thí nghiệm Y-âng về giao thoa ánh sáng, hai khe cách nhau 0.5 mm, màn quan sát cách mặt phẳng chứa hai khe một khoảng D và có thể thay đổi được. Chiếu sáng hai khe bằng ánh sáng đơn sắc có bước sóng $\lambda$ (380 nm $\leq \lambda \leq$ 640 nm). M và N là hai điểm trên màn cách vị trí vân sáng trung tâm lần lượt là 6.4 mm và 9.6 mm. Ban đầu, khi $D = D_1 = 0.8~ m$ thì tại M và N là vị trí của các vân sáng. Khi $D = D_2 = 1.6~m$ thì một trong hai vị trí của M và N là vị trí của vân tối. Tịnh tiến màn từ từ dọc theo phương vuông góc với mặt phẳng chứa hai khe và ra xa hai khe từ vị trí cách hai khe một đoạn $D_1$ đến vị trí cách hai khe một đoạn $D_2$. Troạng quá trình dịch chuyển màn, sồ lần tại N là vị trí của vân sáng (không tính thời điểm ban đầu) là
	
\textcolor{blue}{\textbf{Question}}: In the Young's double-slit experiment on light interference, two slits are separated by 0.5 mm, and the observation screen is at a distance D which can be varied from the plane containing the two slits. The two slits are illuminated by monochromatic light with a wavelength of $\lambda$ (380 nm $\leq \lambda \leq$ 640 nm). M and N are two points on the screen located at a distance of 6.4 mm and 9.6 mm, respectively, from the central bright fringe when $D = D_1 = 0.8~ m$. When $D = D_2 = 1.6~m$, one of the positions of M and N is the position of a dark fringe. The screen is moved slowly along a direction perpendicular to the plane containing the two slits and away from the slits by a distance from $D_1$ to $D_2$. During this displacement, the number of times that N is the position of a bright fringe (excluding the initial position) is

A. 4	

B. 3	

C. 5	

D. 7

\end{mdframed}

\subsubsection{High application level question}

Last but not least, in order to answer the question at the high application level, extensive reasoning is needed.

\begin{mdframed}[linewidth=1pt,linecolor=red]
	
\textcolor{red}{\textbf{Câu hỏi}}: Đặt điện áp $u=120cos\left(100\pi t-\frac{\pi}{6}\right)\left(V\right)$ vào hai đầu đoạn mạch AB mắc nối tiếp gồm: tụ điện có điện dung C thay đổi được; cuộn dây có độ tự cảm L và điện trở r ; điện trở R với R = 2r như hình bên.

\includegraphics[width=0.75\textwidth]{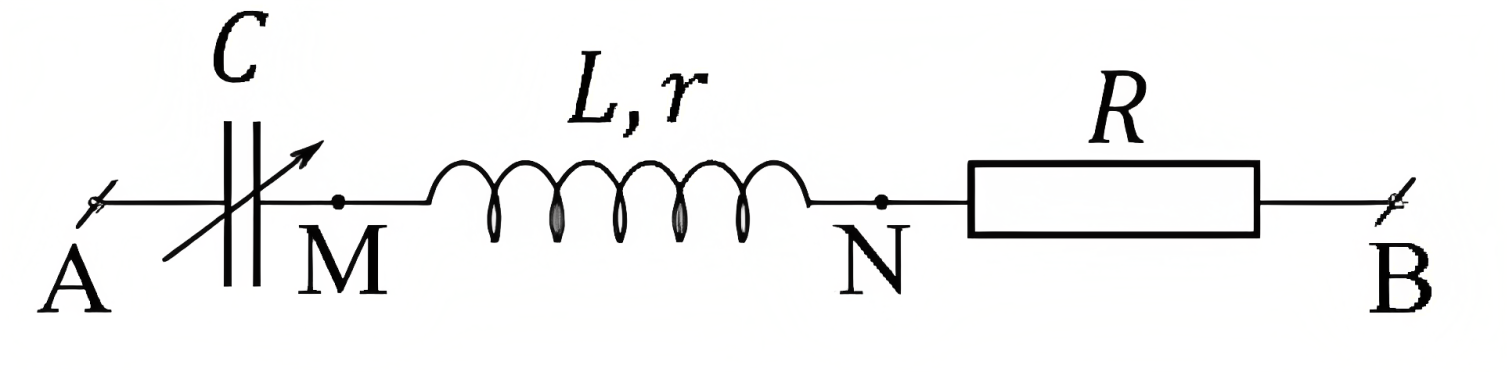}

Khi $C = C_0$ thì điện áp hiệu dụng giữa hai đầu đoạn mạch AN đạt cực tiểu. Khi $C =\frac{C_0}{4}$ thì điện áp hiệu dụng giữa hai đầu đoạn mạch AM đạt cực đại và điện áp giữa hai đầu đoạn mạch MN là $u_MN$. Biểu thức $u_MN$ là

\textcolor{blue}{\textbf{Question}}: Applying a voltage $u=120cos\left(100\pi t-\frac{\pi}{6}\right)\left(V\right)$  to the two ends of circuit AB in series. The capacitor C can be varied, and the circuit also includes an inductor with self-inductance L, a resistor with resistance r, and a resistor R with R = 2r as shown in the figure.

\includegraphics[width=0.75\textwidth]{fig.png}

When $C = C_0$, the RMS voltage between the two ends of the AN segment reaches its minimum value. When $C =\frac{C_0}{4}$, the RMS voltage between the two ends of the AM segment reaches its maximum value and the voltage between the two ends of the MN segment is $u_MN$. The expression for $u_{MN}$ is:

A. $u_{MN}=40\cos{\left(100\pi\ t+\frac{2\pi}{2}\right)}\left(V\right)$

B. $u_{MN}=40\sqrt3\cos{\left(100\pi\ t+\frac{2\pi}{2}\right)}\left(V\right)$

C. $u_{MN}=40\sqrt3\cos{\left(100\pi\ t+\frac{2\pi}{2}\right)}\left(V\right)$

D. $u_{MN}=\ 40\cos{\left(100\pi\ t+\frac{\pi}{2}\right)}\left(V\right)$

\end{mdframed}

\section{Experiments}

We evaluated a evaluation set of VNHSGE dataset (Vietnamese version)~\cite{xuan2023vnhsge} comprised of five illustrative exams (200 multiple-choice questions) offered by the Vietnamese Ministry of Education and Training for the years 2019–2023. The evaluation findings for ChatGPT (Feb 13 version) and BingChat (April 15 version) on this subdataset are presented in this section.

\subsection{ChatGPT and BingChat Responses}

This section introduces ChatGPT and BingChat's reactions to the earlier samples. 

\subsubsection{Knowledge level question}
First, both ChatGPT and BingChat easily solve the knowledge question. The answer is only a formula. 

\begin{mdframed}[linewidth=1pt,linecolor=red]
	
\textcolor{red}{\textbf{Lời giải}}: Trong mạch chỉ có cuộn cảm thuần cho cảm kháng là $Z_L$ thì cường độ dòng điện trong mạch là: $I=\frac{U}{Z_L}$. Chọn C. 
	
\textcolor{blue}{\textbf{Solution}}: The formula for calculating the effective current I in a circuit is $I=\frac{U}{Z_L}$. Choose C. 

\textbf{ChatGPT}: Đáp án là: C. I = U/ZL. Giải thích:
Trong đoạn mạch chỉ có cuộn cảm thuần thì cảm kháng ZL được tính bằng công thức $ZL = \omega L$, trong đó $\omega$ là tần số góc của điện áp xoay chiều. Cường độ dòng điện hiệu dụng I trong đoạn mạch được tính bằng công thức I = U/ZL, trong đó U là giá trị hiệu dụng của điện áp xoay chiều.

C. $I = U/Z_L$. In a circuit with only a pure inductor, the inductive reactance ZL is calculated by the formula $ZL = \omega L$, where $\omega$ is the angular frequency of the AC voltage. The root-mean-square (RMS) current I in the circuit is calculated by the formula $I = U/Z_L$, where U is the RMS value of the AC voltage.

\textbf{BingChat}: Cường độ dòng điện hiệu dụng I trong một đoạn mạch có cảm kháng ZL và điện áp hiệu dụng U được tính bằng công thức $I = \frac{U}{Z_{L}}$. Vì vậy, đáp án đúng là C. $I = \frac{U}{Z_{L}}$.

The effective current intensity I in a circuit segment with inductive impedance ZL and effective voltage U is calculated using the formula $I=\frac{U}{Z_L}$. Therefore, the correct answer is C. $I=\frac{U}{Z_L}$.

\end{mdframed}

\subsubsection{Comprehension level question}

The next case shows that ChatGPT and BingChat find the correct answer although the answer requires reasoning to find the correct answer. 

\begin{mdframed}[linewidth=1pt,linecolor=red]
	
	\textcolor{red}{\textbf{Lời giải}}: Hai dao động cùng pha thì $\varphi_2-\varphi_1=2n\pi$  với $n=\pm0,\pm1,\pm2,\ldots$. Chọn B.   
	
\textcolor{blue}{\textbf{Solution}}: These oscillations are in phase when $\varphi_2-\varphi_1=2n\pi$  with $n=\pm0,\pm1,\pm2,\ldots$. Choose B. 

\textbf{ChatGPT}: 

Đáp án là:
B. $\varphi_1-\varphi_2=2n$ for $k=\pm0,\pm1,\pm2,\ldots$. Giải thích:
Hai dao động điều hòa cùng tần số $A\cos{\left(\omega t+\varphi_1\right)}$ và $B\cos{\left(\omega t+\varphi_2\right)}$ sẽ cùng pha khi chúng đạt giá trị cực đại hoặc giá trị cực tiểu đồng thời. Vì vậy, để hai dao động cùng pha, phải có điều kiện:
$\omega t+\varphi_1=\omega t+\varphi_2+2n$
Tương đương với:
$\varphi_1-\varphi_2=2n \pi$
Do đó, đáp án là B.

B. $\varphi_1-\varphi_2=2n$ for $k=\pm0,\pm1,\pm2,\ldots$. Two harmonic oscillators with the same frequency $A\cos{\left(\omega t+\varphi_1\right)}$ and $B\cos{\left(\omega t+\varphi_2\right)}$ will be in phase when they reach their maximum or minimum values at the same time. Therefore, in order for the two oscillators to be in phase, the following condition must be met: $\omega t+\varphi_1=\omega t+\varphi_2+2n$. Equivalently: $\varphi_1-\varphi_2=2n \pi$. Hence, the answer is B.

\textbf{BingChat}: 

Hai dao động điều hòa cùng tần số có pha ban đầu là $\varphi_{1}$ và $\varphi_{2}$ cùng pha khi hiệu pha ban đầu của chúng là bội số của $2\pi$. Vì vậy, $\varphi_{1} - \varphi_{2} = 2n\pi$ với n là số nguyên. Đáp án đúng là B. $\varphi_{1} - \varphi_{2} = 2n\pi$ với $n=\pm0,\pm1,\pm2,\ldots$

Two harmonic oscillations of the same frequency with initial phases of $\varphi_1$ and $\varphi_2$ are in phase when the initial phase difference between them is a multiple of $2\pi$. Therefore, $\varphi_1-\varphi_2=2n \pi$ with n being an integer. The correct answer is B. $\varphi_1-\varphi_2=2n \pi$ with $n=\pm0,\pm1,\pm2,\ldots$

\end{mdframed}

\subsubsection{Application level question}

Both ChatGPT and BingChat were unable to come up with an answer for the application-level question, which necessitates comprehensive knowledge for determining the right response. The question requires multiple steps in reasoning to find the correct answer. Neither ChatGPT nor BingChat can provide a reasonable solution. 

\begin{mdframed}[linewidth=1pt,linecolor=red]
	
	\textcolor{red}{\textbf{Lời giải}}:
	 
	 Khi $D=D_1$, khoảng vân giao thoa là: $i_1=\frac{\lambda D_1}{a}$
	 Tại $\mathrm{M}$ và $\mathrm{N}$ ta có:
	 $$
	 \begin{aligned}
	 	& \frac{k_N}{k_M}=\frac{x_N}{x_M}=\frac{9,6}{6,4}=\frac{3}{2} \\
	 	& \Rightarrow\left\{\begin{array}{l}
	 		k_N=3 p \\
	 		k_M=2 p
	 	\end{array}\right.
	 \end{aligned}
	 $$
	 Lại có: $x=k_1 i_1$
	 
	 Khi $D=D_2$, khoảng vân là:
	 $$
	 i_2=\frac{\lambda D_2}{a}=\frac{\lambda \cdot 2 D_1}{a}=2 i_1
	 $$
	 Tại một vị trí cho vân tối, ta có:
	 $$
	 x=k_2 i_2=k_1 i_1 \Rightarrow k_2 \cdot 2 i_1=k_1 i_1 \Rightarrow k_1=2 k_2
	 $$
	 Mà $k_2$ là số bán nguyên $\rightarrow k_1$ là số lẻ
	 $\rightarrow$ tại $\mathrm{N}$ thỏa mãn điều kiện đề bài
	 Mặt khác:
	 $$
	 x_M=k_1 i_1 \Rightarrow 9,6 \cdot 10^{-3}=k_1 \cdot \frac{\lambda \cdot 0,8}{0,5.10^{-3}} \Rightarrow k_1 \lambda=6.10^{-6}
	 $$
	
	Theo đề bài ta có:
	$$
	\begin{aligned}
		& 0,38 \cdot 10^{-6} \leq \lambda \leq 0,64 \cdot 10^{-6} \\
		& \Rightarrow 0,38 \cdot 10^{-6} \leq \frac{6 \cdot 10^{-6}}{k_1} \leq 0,64 \cdot 10^{-6} \\
		& \Rightarrow 15,8 \geq k_1 \geq 9,3 \Rightarrow k_1=15 \\
		& \Rightarrow \lambda=0,4 \cdot 10^{-6}(\mathrm{~m})
	\end{aligned}
	$$
	Khi $D=D_2$ ta có:
	$$
	\begin{aligned}
		& x_N=k_2 i_2 \Rightarrow 9,6 \cdot 10^{-3}=k_2 \cdot \frac{0,4 \cdot 10^{-6} \cdot 1,6}{0,5 \cdot 10^{-3}} \\
		& \Rightarrow k_2=7,5
	\end{aligned}
	$$
	Vậy có 7 lần tại $\mathrm{N}$ có vân sáng, ứng với $\mathrm{k}=8 ; 9 ; 10 ; 11 ; 12 ; 13 ; 14$. Chọn D. 
	
\textcolor{blue}{\textbf{Solution}}: 

	 When $D=D_1$, the fringe spacing is given by: $i_1=\frac{\lambda D_1}{a}$
At points $\mathrm{M}$ and $\mathrm{N}$, we have:
$$
\begin{aligned}
	& \frac{k_N}{k_M}=\frac{x_N}{x_M}=\frac{9,6}{6,4}=\frac{3}{2} \\
	& \Rightarrow\left\{\begin{array}{l}
		k_N=3 p \\
		k_M=2 p
	\end{array}\right.
\end{aligned}
$$
Also, we have: $x=k_1 i_1$

When $D=D_2$, the fringe spacing is:
$$
i_2=\frac{\lambda D_2}{a}=\frac{\lambda \cdot 2 D_1}{a}=2 i_1
$$
At a position of a dark fringe, we have:
$$
x=k_2 i_2=k_1 i_1 \Rightarrow k_2 \cdot 2 i_1=k_1 i_1 \Rightarrow k_1=2 k_2
$$
Since $k_2$ is an integer, $k_1$ is odd. Therefore, point $\mathrm{N}$ satisfies the given condition.
Furthermore:
$$
x_M=k_1 i_1 \Rightarrow 9,6 \cdot 10^{-3}=k_1 \cdot \frac{\lambda \cdot 0,8}{0,5.10^{-3}} \Rightarrow k_1 \lambda=6.10^{-6}
$$

According to the problem statement:
$$
\begin{aligned}
	& 0,38 \cdot 10^{-6} \leq \lambda \leq 0,64 \cdot 10^{-6} \\
	& \Rightarrow 0,38 \cdot 10^{-6} \leq \frac{6 \cdot 10^{-6}}{k_1} \leq 0,64 \cdot 10^{-6} \\
	& \Rightarrow 15,8 \geq k_1 \geq 9,3 \Rightarrow k_1=15 \\
	& \Rightarrow \lambda=0,4 \cdot 10^{-6}(\mathrm{~m})
\end{aligned}
$$
When $D=D_2$, we have:
$$
\begin{aligned}
	& x_N=k_2 i_2 \Rightarrow 9,6 \cdot 10^{-3}=k_2 \cdot \frac{0,4 \cdot 10^{-6} \cdot 1,6}{0,5 \cdot 10^{-3}} \\
	& \Rightarrow k_2=7,5
\end{aligned}
$$
Thus, there are 7 instances of bright fringes at point $\mathrm{N}$, corresponding to $\mathrm{k}=8, 9, 10, 11, 12, 13, 14$. Choose D.

\textbf{ChatGPT}: Theo công thức về khoảng cách giữa hai vân sáng: $d\sin\theta=m\lambda$ trong đó, m là số thứ tự của vân sáng, d là khoảng cách giữa hai khe, $\lambda$ là bước sóng của ánh sáng và $\theta$ là góc giữa đường thẳng nối điểm chiếu và trung điểm hai khe với đường thẳng vuông góc với mặt phẳng hai khe. Ta có thể tính được khoảng cách giữa hai vân sáng liên tiếp tại N khi màn dịch chuyển một khoảng $\Delta D$ bằng công thức: $\Delta d=\frac{d\Delta D}{D_1D_2}$ Do đó, số lần tại $N$ là vị trí của vân sáng sẽ là: $n=\left\lfloor\frac{\Delta D}{\frac{\Delta d}{\sin\theta}}\right\rfloor$
Với $D=D_1=0,8 m$, ta tính được: $\sin\theta=\frac{Mn-MN}{D_1}=\frac{6,4\times 10^{-9}}{0,8}=8\times 10^{-9}$ Dùng công thức về khoảng cách giữa hai vân sáng, ta tính được bước sóng của ánh sáng là: $\lambda=\frac{d\sin\theta}{m}=\frac{0,5\times 10^{-3}\times 8\times 10^{-9}}{1}=4\times 10^{-12} m$
Với $D=D_2=1,6 m$, ta tính được: $\sin\theta=\frac{Nn-MN}{D_2}=\frac{9,6\times 10^{-9}}{1,6}=6\times 10^{-9}$ Dùng công thức về khoảng cách giữa hai vân sáng, ta tìm được số thứ tự của vân tối tại N là: $m=\frac{d\sin\theta}{\lambda}=\frac{0,5\times 10^{-3}\times 6\times 10^{-9}}{4\times 10^{-12}}=750$
Do đó, nếu ta dịch chuyển màn dọc theo phương vuông góc với mặt phẳng hai khe từ $D_1$ đến $D_2$, thì sẽ có 750-1=749 vân sáng liên tiếp tại N, tương ứng với số lần tại N là vị trí của vân sáng là 4.
Vậy đáp án đúng là A. 4.

According to the formula for the distance between two bright fringes: $d \sin \theta=m$, where $\mathrm{m}$ is the order number of the bright fringe, $\mathrm{d}$ is the distance between two slits, lambda is the wavelength of light, and theta is the angle between the line connecting the point of projection and the midpoint of the two slits with the line perpendicular to the plane of the two slits. We can calculate the distance between two consecutive bright fringes at $\mathrm{N}$ when the screen is shifted by a distance $\Delta D$ using the formula: $\Delta d=\frac{d \Delta D}{D_1 D_2}$. Therefore, the number of times the bright fringe appears at $\mathrm{N}$ is given by: $n=\mid \frac{\Delta D}{\frac{\Delta d}{\sin \theta}}$. For $D=D_1=0.8 \mathrm{~m}$, we can calculate: $\sin \theta=\frac{M n-M N}{D_1}=\frac{6.4 \times 10^{-9}}{0.8}=8 \times 10^{-9}$. Using the formula for the distance between two bright fringes, we can calculate the wavelength of light: $\lambda=\frac{d \sin \theta}{m}=\frac{0.5 \times 10^{-3} \times 8 \times 10^{-9}}{1}=4 \times 10^{-12} \mathrm{~m}$. For $D=D_2=$ $1.6 \mathrm{~m}$, we can calculate: $\sin \theta=\frac{N n-M N}{D_2}=\frac{9.6 \times 10^{-9}}{1.6}=6 \times 10^{-9}$. Using the formula for the distance between two bright fringes, we can find the order number of the dark fringe at $\mathrm{N}: m=\frac{d \sin \theta}{\lambda}=\frac{0.5 \times 10^{-3} \times 6 \times 10^{-9}}{4 \times 10^{-12}}=$ 750.Therefore, if we shift the screen perpendicular to the plane of the two slits from $D_1$ to $D_2$, there will be $750-1=749$ consecutive bright fringes at $\mathrm{N}$, corresponding to the position of the bright fringe at $\mathrm{N}$ being 4. $=>\mathbf{A}$

\textbf{BingChat}: 

Câu trả lời đúng là B. 3. Khoảng cách giữa hai vân sáng liên tiếp trên màn quan sát trong thí nghiệm Y-âng về giao thoa ánh sáng có thể được tính bằng công thức: $\Delta y = \frac{\lambda D}{d}$. Do đó, khi $D = D1 = 0,8 m$ thì khoảng cách giữa hai vân sáng liên tiếp trên màn quan sát là $\Delta y1 = \frac{\lambda D1}{d} = \frac{0,8\lambda}{0,5\times 10^{-3}} (m)$. Khi $D = D2 = 1,6 m$ thì khoảng cách giữa hai vân sáng liên tiếp trên màn quan sát là $\Delta y2 = \frac{\lambda D2}{d} = \frac{1,6\lambda}{0,5\times 10^{-3}} (m)$. Vì tại N là vị trí của vân sáng khi D = D1 và là vị trí của vân tối khi D = D2 nên ta có: $9,6\times 10^{-3} = k\Delta y1 và 9,6\times 10^{-3} + \frac{1}{2}\Delta y2 = k’\Delta y2$. Từ đó ta có: $k’ - k = \frac{9,6\times 10^{-3}}{\Delta y2} - \frac{9,6\times 10^{-3}}{\Delta y1} + \frac{1}{2} = 3$. Do đó, trong quá trình dịch chuyển màn từ vị trí cách hai khe một đoạn D1 đến vị trí cách hai khe một đoạn D2 thì số lần tại N là vị trí của vân sáng (không tính thời điểm ban đầu) là 3.

The correct answer is B. 3. The distance between two consecutive bright fringes on the observation screen in the Y-âng experiment on light interference can be calculated using the formula: $\Delta y=\frac{\lambda D}{d}$. Therefore, when $\mathrm{D}=\mathrm{D}_1=0.8 \mathrm{~m}$, the distance between two consecutive bright fringes on the observation screen is $\Delta y 1=\frac{\lambda D 1}{d}=$ $\frac{0.8 \lambda}{0.5 \times 10^{-3}}(\mathrm{~m})$. When $\mathrm{D}=\mathrm{D}_2=1.6 \mathrm{~m}$, the distance between two consecutive bright fringes on the observation screen is $\Delta y 2=\frac{\lambda D 2}{d}=\frac{1.6 \lambda}{0.5 \times 10^{-3}}(\mathrm{~m})$. Since at $\mathrm{N}$ is the position of the bright fringe when $\mathrm{D}=\mathrm{D}_1$ and is the position of the dark fringe when $\mathrm{D}=\mathrm{D}_2$, we have: $9.6 \times 10^{-3}=$ $k \Delta y 1$ and $9.6 \times 10^{-3}+\frac{1}{2} \Delta y 2=k^{\prime} \Delta y 2$. From this we have: $k^{\prime}-k=$ $\frac{9.6 \times 10^{-3}}{\Delta y 2}-\frac{9.6 \times 10^{-3}}{\Delta y 1}+\frac{1}{2}=3$. Therefore, during the process of moving the screen from a position a distance $\mathrm{D}_1$ from the two slits to a position a distance D2 from the two slits, the number of times at $\mathrm{N}$ is the position of a bright fringe (not counting the initial time) is 3 .

\end{mdframed}

\subsubsection{High application level question}

BingChat doesn't give an answer for questions that required information from images, while ChatGPT proposes a solution. Although the question is at the high application level, ChatGPT's answer is correct.  This is a special case because both ChatGPT and BingChat are usually unable to answer questions at the high application level.

\begin{mdframed}[linewidth=1pt,linecolor=red]
	
	\textcolor{red}{\textbf{Lời giải}}:

Khi $C=C_0 \Rightarrow Z_C=Z_{C_0}$, điện áp hiệu dụng giữa hai đầu đoạn mạch $\mathrm{AN}$ đạt cực tiểu, ta có:
$$
\begin{aligned}
	& U_{A N}=\frac{U \cdot \sqrt{r^2+\left(Z_L-Z_{C_0}\right)^2}}{\sqrt{(R+r)^2+\left(Z_L-Z_{C_0}\right)^2}}=\frac{U \sqrt{r^2+Z_{L C}^2}}{\sqrt{(R+r)^2+Z_{L C}^2}} \\
	& \Rightarrow U_{A N}=\frac{U}{\sqrt{\frac{R^2+2 R r}{r^2+Z_{L C}}+1}}
\end{aligned}
$$

Để $U_{A N \min } \Leftrightarrow\left(\frac{R^2+2 R r}{r^2+Z_{L C}}\right)_{\max } \Leftrightarrow Z_{L C \min }=0 \Rightarrow Z_L=Z_{C_0}$

Đặt $Z_{C_0}=1 \Rightarrow Z_L=1$

Khi $C=\frac{C_0}{4} \Rightarrow Z_C=4 Z_{C_0}=4$

Điện áp hiệu dụng $U_{A M \max } \Leftrightarrow Z_C=\frac{(R+r)^2+Z_L^2}{Z_L}$
$$
\begin{aligned}
	& \Rightarrow 4=\frac{(2 r+r)^2+1^2}{1} \Rightarrow r=\frac{1}{\sqrt{3}} \\
	& \Rightarrow R=2 r=\frac{2}{\sqrt{3}}
\end{aligned}
$$

Độ lệch pha giữa điện áp hiệu dụng hai đầu đoạn mạch $\mathrm{AB}$ và cường độ dòng điện là:
$$
\begin{aligned}
	& \tan \varphi=\frac{Z_L-Z_C}{R+r}=\frac{1-4}{\frac{2}{\sqrt{3}}+\frac{1}{\sqrt{3}}}=-\sqrt{3} \Rightarrow \varphi=-\frac{\pi}{3} \\
	& \varphi_u-\varphi_i=-\frac{\pi}{3} \Rightarrow \varphi_i=\varphi_u+\frac{\pi}{3}=-\frac{\pi}{6}+\frac{\pi}{3}=\frac{\pi}{6}(\mathrm{rad})
\end{aligned}
$$

Độ lệch pha giữa điện áp hai đầu đoạn mạch $\mathrm{MN}$ và cường độ dòng điện là:

$$
\begin{aligned}
	& \tan \varphi_{M N}=\frac{Z_L}{r}=\frac{1}{\frac{1}{\sqrt{3}}}=\sqrt{3}=\tan \frac{\pi}{3} \\
	& \Rightarrow \varphi_{u_{M N}}-\varphi_i=\frac{\pi}{3} \Rightarrow \varphi_{u_{M N}}=\varphi_i+\frac{\pi}{3}=\frac{\pi}{2}(\mathrm{rad})
\end{aligned}
$$
Lại có:
$$
\begin{aligned}
	& \frac{U_{0 M N}}{U_{0 A B}}=\frac{\sqrt{r^2+Z_L^2}}{\sqrt{(R+r)^2+\left(Z_L-Z_C\right)^2}} \Rightarrow \frac{U_{0 M N}}{120}=\frac{1}{3} \\
	& \Rightarrow U_{0 M N}=40(V) \\
	& \Rightarrow u_{M N}=40 \cos \left(100 \pi t+\frac{\pi}{2}\right)(V)
\end{aligned}
$$
Chọn D.

\textcolor{blue}{\textbf{Solution}}:

When $C=C_0 \Rightarrow Z_C=Z_{C_0}$, the effective voltage across the circuit segment $\mathrm{AN}$ reaches a minimum when:
$$
\begin{aligned}
	& U_{A N}=\frac{U \cdot \sqrt{r^2+\left(Z_L-Z_{C_0}\right)^2}}{\sqrt{(R+r)^2+\left(Z_L-Z_{C_0}\right)^2}}=\frac{U \sqrt{r^2+Z_{L C}^2}}{\sqrt{(R+r)^2+Z_{L C}^2}} \\
	& \Rightarrow U_{A N}=\frac{U}{\sqrt{\frac{R^2+2 R r}{r^2+Z_{L C}}+1}}
\end{aligned}
$$

For $U_{A N \min } \Leftrightarrow\left(\frac{R^2+2 R r}{r^2+Z_{L C}}\right)_{\max } \Leftrightarrow Z_{L C \min }=0 \Rightarrow Z_L=Z_{C_0}$

Taking $Z_{C_0}=1 \Rightarrow Z_L=1$

When $C=\frac{C_0}{4} \Rightarrow Z_C=4 Z_{C_0}=4$

The effective voltage $U_{A M \max } \Leftrightarrow Z_C=\frac{(R+r)^2+Z_L^2}{Z_L}$
$$
\begin{aligned}
	& \Rightarrow 4=\frac{(2 r+r)^2+1^2}{1} \Rightarrow r=\frac{1}{\sqrt{3}} \\
	& \Rightarrow R=2 r=\frac{2}{\sqrt{3}}
\end{aligned}
$$

The phase difference between the effective voltage across segment $\mathrm{AB}$ and the current intensity is:
$$
\begin{aligned}
	& \tan \varphi=\frac{Z_L-Z_C}{R+r}=\frac{1-4}{\frac{2}{\sqrt{3}}+\frac{1}{\sqrt{3}}}=-\sqrt{3} \Rightarrow \varphi=-\frac{\pi}{3} \\
	& \varphi_u-\varphi_i=-\frac{\pi}{3} \Rightarrow \varphi_i=\varphi_u+\frac{\pi}{3}=-\frac{\pi}{6}+\frac{\pi}{3}=\frac{\pi}{6}(\mathrm{rad})
\end{aligned}
$$

The phase difference between the voltage across segment $\mathrm{MN}$ and the current intensity is:

$$
\begin{aligned}
	& \tan \varphi_{M N}=\frac{Z_L}{r}=\frac{1}{\frac{1}{\sqrt{3}}}=\sqrt{3}=\tan \frac{\pi}{3} \\
	& \Rightarrow \varphi_{u_{M N}}-\varphi_i=\frac{\pi}{3} \Rightarrow \varphi_{u_{M N}}=\varphi_i+\frac{\pi}{3}=\frac{\pi}{2}(\mathrm{rad})
\end{aligned}
$$

Also, we have:
$$
\begin{aligned}
	& \frac{U_{0 M N}}{U_{0 A B}}=\frac{\sqrt{r^2+Z_L^2}}{\sqrt{(R+r)^2+\left(Z_L-Z_C\right)^2}} \Rightarrow \frac{U_{0 M N}}{120}=\frac{1}{3} \\
	& \Rightarrow U_{0 M N}=40(V) \\
	& \Rightarrow u_{M N}=40 \cos \left(100 \pi t+\frac{\pi}{2}\right)(V)
\end{aligned}
$$

Therefore, the correct answer is D.

\textbf{ChatGPT}: 

Ta dùng công thức tổng quát để tính điện áp hiệu dụng trên mạch C-L-r: 

$$U_{AN} = U \sqrt{\dfrac{R^2+(2\pi L - \dfrac{1}{2\pi C})^2}{R^2+(2\pi L)^2}}$$ $$= 120 \sqrt{\dfrac{(2r)^2+(2\pi L - \dfrac{1}{2\pi C_0})^2}{(2r)^2+(2\pi L)^2}}$$ 

Ta có: $\dfrac{1}{2\pi C_0} = 2r \Rightarrow C_0 = \dfrac{1}{4\pi^2r^2}$. Thay giá trị $C = \dfrac{C_0}{4} = \dfrac{1}{16\pi^2r^2}$ vào công thức, ta được: $U_{AM} = U_{C} = 120\sqrt{3}$

Điện áp trên mạch L-r khi đó là: $$U_{MN} = U_{AN} - U_C = 40 \cos \left(100\pi t + \dfrac{\pi}{2} \right) \mathrm{V}$$ Vậy đáp án chính xác là $\boxed{\textbf{(D)}}$.

We use the general formula to calculate the effective voltage on the C-L-r circuit:

$$U_{AN} = U \sqrt{\dfrac{R^2+(2\pi L - \dfrac{1}{2\pi C})^2}{R^2+(2\pi L)^2}}$$ $$= 120 \sqrt{\dfrac{(2r)^2+(2\pi L - \dfrac{1}{2\pi C_0})^2}{(2r)^2+(2\pi L)^2}}$$ 

We have: $\dfrac{1}{2\pi C_0} = 2r \Rightarrow C_0 = \dfrac{1}{4\pi^2r^2}$. Substituting the value of $C = \dfrac{C_0}{4} = \dfrac{1}{16\pi^2r^2}$ into the formula, we get: $U_{A M}=U_C=120 \sqrt{3}$. 

The voltage on the L-r circuit is then: 
$$U_{M N}=U_{A N}-U_C=40 \cos (100 \pi t+\pi / 2) \mathrm{V}$$
So the correct answer is $\mathbf{D}$.

\textbf{BingChat}: Câu hỏi của bạn là một câu hỏi về mạch điện xoay chiều và điện áp hiệu dụng. Để giải quyết câu hỏi này, chúng ta cần sử dụng các nguyên lý về mạch điện xoay chiều để tính toán điện áp giữa hai đầu đoạn mạch MN.

Tuy nhiên, để giải quyết câu hỏi này chúng ta cần biết thêm thông tin về các giá trị của các thông số như độ tự cảm L và điện trở r của cuộn dây. Bạn có thể cung cấp thêm thông tin để tôi có thể giúp đỡ bạn tốt hơn không?

Your question is about alternating current circuits and effective voltage. To solve this question, we need to use the principles of alternating current circuits to calculate the voltage between the two ends of the MN circuit segment. However, to solve this question we need to know more information about the values of parameters such as inductance L and resistance r of the coil. Can you provide more information so I can help you better?

\end{mdframed}

\subsection{Performance}

\subsubsection{Question order}

The difficulty of the questions determines how they are arranged. The questions are divided into three categories: knowledge level (questions 1–20), comprehension level (questions 20–30), and application and high application levels (questions 30–40). The results from ChatGPT and BingChat are shown in Table~\ref{tabl:accuracy} in the order of the questions. We also suggest two values: Max and Min. Max reflects the capability of ChatGPT and BingChat to answer questions correctly in the greatest possible way. The worst-case scenario for ChatGPT and BingChat, or the models' capability to give wrong replies, is represented by Min (1-Min).

$$
	 \mathrm{Max}=\left\{\begin{array}{lc}
		0 & \text{ if ChatGPT }=0 \text{ and BingChat }=0 \\
		1 & \text{ otherwise }
	\end{array}\right. 
$$
$$
\mathrm{Min}= \begin{cases}1 & \text{ if ChatGPT }=1 \text{ and BingChat }=1 \\
	0 & \text{ otherwise }\end{cases}
$$

\begin{figure*}[h!]
	\begin{center}
		\begin{tikzpicture}
			\begin{axis}
				[
				ylabel={Accuracy},
				legend style={at={(0.5,-0.25)}, 	
					anchor=north,legend columns=-1}, 
				symbolic x coords={
					1,
					2,
					3,
					4,
					5,
					6,
					7,
					8,
					9,
					10,
					11,
					12,
					13,
					14,
					15,
					16,
					17,
					18,
					19,
					20,
					21,
					22,
					23,
					24,
					25,
					26,
					27,
					28,
					29,
					30,
					31,
					32,
					33,
					34,
					35,
					36,
					37,
					38,
					39,
					40,
				},
				xtick=data,
				x tick label style={rotate=90,anchor=east},
				ymin=0,
				width=\textwidth, 
				height=4cm, 
				width=16cm,
				axis x line*=bottom, axis y line*=left
				]
				\addplot[color=green,mark=*,semithick, mark options={solid,}]
				coordinates{
					(1,60)
					(2,60)
					(3,80)
					(4,80)
					(5,80)
					(6,100)
					(7,80)
					(8,60)
					(9,60)
					(10,80)
					(11,60)
					(12,100)
					(13,80)
					(14,40)
					(15,100)
					(16,60)
					(17,60)
					(18,60)
					(19,40)
					(20,80)
					(21,60)
					(22,60)
					(23,60)
					(24,80)
					(25,80)
					(26,100)
					(27,60)
					(28,60)
					(29,100)
					(30,40)
					(31,60)
					(32,60)
					(33,20)
					(34,20)
					(35,0)
					(36,40)
					(37,40)
					(38,20)
					(39,20)
					(40,40)
				};
				
				\addplot[color=red,mark=square*,semithick, mark options={solid,}]
				coordinates{
					(1,100)
					(2,80)
					(3,100)
					(4,60)
					(5,100)
					(6,60)
					(7,80)
					(8,80)
					(9,100)
					(10,100)
					(11,80)
					(12,100)
					(13,100)
					(14,80)
					(15,80)
					(16,40)
					(17,80)
					(18,40)
					(19,80)
					(20,100)
					(21,100)
					(22,100)
					(23,20)
					(24,100)
					(25,100)
					(26,80)
					(27,80)
					(28,60)
					(29,100)
					(30,40)
					(31,60)
					(32,0)
					(33,20)
					(34,20)
					(35,40)
					(36,20)
					(37,20)
					(38,0)
					(39,40)
					(40,0)
				};
				
				\addplot[color=gray,mark=triangle*,semithick, mark options={solid,}]
				coordinates{
					(1,60)
					(2,60)
					(3,80)
					(4,60)
					(5,80)
					(6,60)
					(7,80)
					(8,60)
					(9,60)
					(10,80)
					(11,40)
					(12,100)
					(13,80)
					(14,20)
					(15,80)
					(16,20)
					(17,60)
					(18,40)
					(19,40)
					(20,80)
					(21,60)
					(22,60)
					(23,20)
					(24,80)
					(25,80)
					(26,80)
					(27,40)
					(28,60)
					(29,100)
					(30,40)
					(31,60)
					(32,0)
					(33,0)
					(34,0)
					(35,0)
					(36,0)
					(37,20)
					(38,0)
					(39,0)
					(40,0)
				};
				
				\addplot[color=yellow,mark=star,semithick, mark options={solid,}]
				coordinates{
					(1,100)
					(2,80)
					(3,100)
					(4,80)
					(5,100)
					(6,100)
					(7,80)
					(8,80)
					(9,100)
					(10,100)
					(11,100)
					(12,100)
					(13,100)
					(14,100)
					(15,100)
					(16,80)
					(17,80)
					(18,60)
					(19,80)
					(20,100)
					(21,100)
					(22,100)
					(23,60)
					(24,100)
					(25,100)
					(26,100)
					(27,100)
					(28,60)
					(29,100)
					(30,40)
					(31,60)
					(32,60)
					(33,40)
					(34,40)
					(35,40)
					(36,60)
					(37,40)
					(38,20)
					(39,60)
					(40,40)
				};
				\legend{ChatGPT, BingChat, Min, Max}
			\end{axis}
		\end{tikzpicture}
	\end{center}
	\caption{Correctness average of ChatGPT, BingChat, Min and Max in question order.}
	\label{fig:order}
\end{figure*}
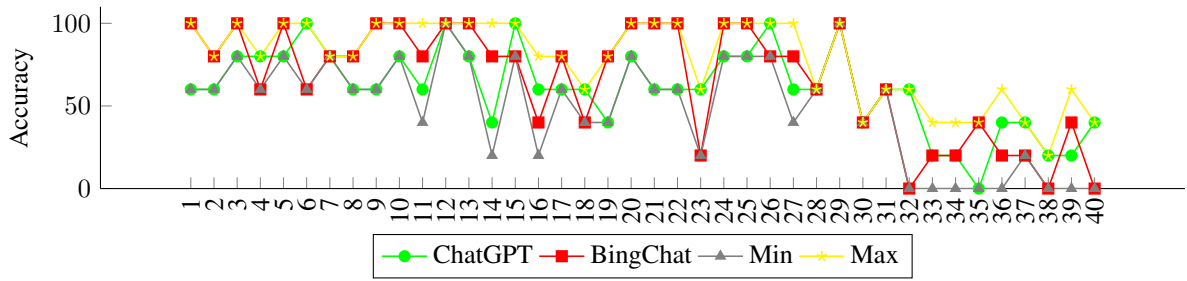

The values Max and Min enable us to evaluate the best and worst LLMs performance on the VNHSGE dataset. For instance, BingChat provided the incorrect response to inquiry "x" while ChatGPT provided the right one, resulting in Max being true (value "1") and Min being false (value "0"). Since the order of the questions affects how difficult they are, we can assess the correctness of the answers based on the order of the questions, allowing us to assess the proficiency of the LLMs based on the difficulty of the questions.

\begingroup
\renewcommand{\arraystretch}{1.5} 
\begin{table}[h!]
	\caption{Accuracy}
	\label{tabl:accuracy}
	\resizebox{0.5\textwidth}{!}{%
	\begin{tabular}{|c|c|c|c|c|}
		\hline
		\textbf{Question Order} & \textbf{ChatGPT} & \textbf{BingChat} & \textbf{Max} & \textbf{Min} \\ \hline
		1                       & 1                & 1                 & 1            & 1            \\ \hline
		2                       & 0                & 1                 & 1            & 0            \\ \hline
		3                       & 1                & 0                 & 1            & 0            \\ \hline
		4                       & 1                & 0                 & 1            & 0            \\ \hline
		5                       & 1                & 1                 & 1            & 1            \\ \hline
		...                     & ...              & ...               & ...          & ...          \\ \hline
	\end{tabular}
	}
\end{table}
\endgroup

Fig.~\ref{fig:order} presents the 5-year average results, which show that ChatGPT has an accuracy rate of over 50$\%$ for questions 1-32 but falls below 50$\%$ for questions 33-40. On the other hand, BingChat, Min, and Max can provide correct answers for questions 1-31, 1-31, and 1-32 with an accuracy of more than 50$\%$, respectively. However, both BingChat and Min show a decrease in accuracy below 50$\%$ for some questions. Notably, Min's accuracy rate drops to almost 0$\%$ from question 32 onwards. Analysis of ChatGPT and BingChat shows that these models can answer questions at the knowledge, comprehension, and application levels, but they face difficulties in solving questions at high application levels. The results suggest that these models need further improvement to perform better in advanced application-level questions.

\subsubsection{Performance evaluation}

Table~\ref{tabl:performance} displays the performance of LLMs for each year and their averages. ChatGPT's performance is $61 \%$ (from $57.5 \%$ to $65 \%$ ), while BingChat obtained $66 \%$ (from $72.5 \%$ to $55 \%$ ). Max performed consistently well, $78.5 \%$, while Min's performance is $48.5 \%$. Interestingly, ChatGPT outperformed BingChat only in 2019. These findings suggest that while each LLM has its strengths and weaknesses, Max is the most consistent performer across all years. However, further investigation is needed to identify the factors that contributed to each model's performance. 

\begingroup
\renewcommand{\arraystretch}{1.5} 
\begin{table}[h!]
	\caption{Performance}
	\label{tabl:performance}
	\resizebox{0.5\textwidth}{!}{%
	\begin{tabular}{c|c|c|c|c|c|}
		\cline{2-6}
		\multicolumn{1}{l|}{}               & \textbf{Max}  & \textbf{ChatGPT} & \textbf{BingChat} & \textbf{Min}  & \textbf{BingChat/ChatGPT}\\ \hline
		\multicolumn{1}{|c|}{\textbf{2019}} & 75            & 60               & 55                & 40            & -8,33                                                                      \\ \hline
		\multicolumn{1}{|c|}{\textbf{2020}} & 80            & 62.5             & 67.5              & 50            & 8,00                                                                       \\ \hline
		\multicolumn{1}{|c|}{\textbf{2021}} & 80            & 60               & 67.5              & 47.5          & 12,50                                                                      \\ \hline
		\multicolumn{1}{|c|}{\textbf{2022}} & 77.5          & 65               & 67.5              & 55            & 3,85                                                                       \\ \hline
		\multicolumn{1}{|c|}{\textbf{2023}} & 80            & 57.5             & 72.5              & 50            & 26,09                                                                      \\ \hline
		\multicolumn{1}{|c|}{\textbf{AVG}}  & \textbf{78.5} & \textbf{61}      & \textbf{66}       & \textbf{48.5} & \textbf{8,42}                                                              \\ \hline
	\end{tabular}
	}
\end{table}
\endgroup


\begin{figure}[h!]
	\begin{center}
		\begin{tikzpicture}
			\pgfplotstableread[col sep=comma]{performance}\csvdata
			\pgfplotstabletranspose\datatransposed{\csvdata} 
			\begin{axis}[
				boxplot/draw direction = y,
				x axis line style = {opacity=0},
				axis x line* = bottom,
				axis y line = left,
				enlarge y limits,
				ymajorgrids,
				xtick = {1, 2, 3, 4},
				xticklabel style = {align=center, font=\small, rotate=0},
				xticklabels = {Max, ChatGPT, BingChat, Min},
				xtick style = {draw=none}, 
				ylabel = {Performance (\%)},
				ytick = {40, 80}
				]
				\foreach \n in {1,...,4} {
					\addplot+[boxplot, fill, draw=black] table[y index=\n] {\datatransposed};
				}
			\end{axis}
		\end{tikzpicture}
		\caption{Stability of LLMs response.}
		\label{fig:performance}
	\end{center}
\end{figure}
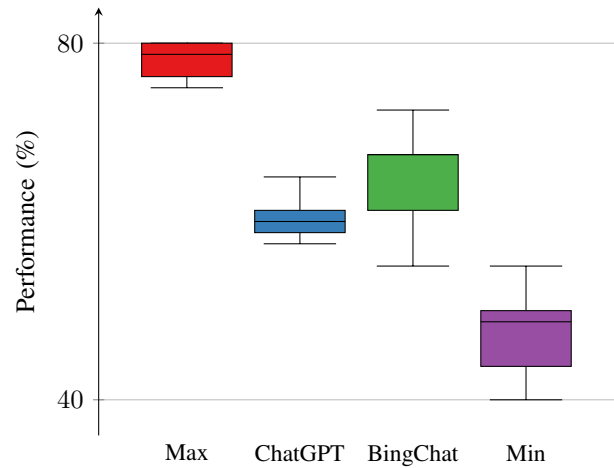	

In Fig.~\ref{fig:performance}, the consistency of responses given by ChatGPT and BingChat on the VNHSGE dataset is demonstrated. The results show that ChatGPT is more consistent than BingChat. This information can help in understanding the strengths and weaknesses of different LLMs, which can be used to guide their use in various applications. Furthermore, the observed differences in consistency between the two models may have practical implications for the reliability of their responses. For instance, it is important to consider the level of consistency in determining the trustworthiness of AI-generated responses, especially in contexts where errors or inaccuracies can have significant consequences. Further research could investigate the reasons for the observed differences in consistency and explore ways to improve the reliability of AI-generated responses.

\subsubsection{Comparison to other exams} 

\begin{figure}[h!]
	\begin{center}
		\begin{tikzpicture}
			\begin{axis}[
				symbolic y coords={
					Max-VNHSGE Physics,
					GPT-4-AP Physics 2,
					BingChat-VNHSGE Physics,
					ChatGPT-VNHSGE Physics,
					Min-VNHSGE Physics,
					GPT-$3.5$-AP Physics 2,
				},
				ytick=data,
				xbar,
				bar width=3pt,
				xmin=0,
				enlarge y limits,
				height=4cm, 
				width=0.325\textwidth,
				axis x line*=bottom, axis y line*=left
				]
				
				\addplot coordinates {
					(78.5,Max-VNHSGE Physics)
					(66,GPT-4-AP Physics 2)
					(66,BingChat-VNHSGE Physics)
					(61,ChatGPT-VNHSGE Physics)
					(48.5,Min-VNHSGE Physics)
					(30,GPT-$3.5$-AP Physics 2)
				};
			\end{axis}
		\end{tikzpicture}
		\caption{Performance on VNHSGE dataset and other exams.}
		\label{fig:other_dataset}
	\end{center}
\end{figure}
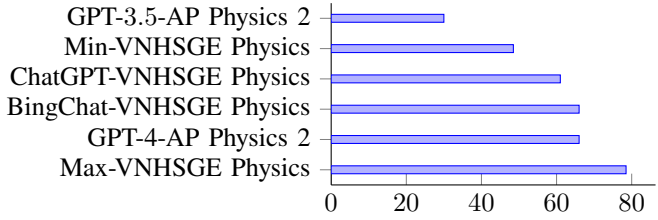

Fig.~\ref{fig:other_dataset} presents a comparison between the performances of ChatGPT and BingChat on the VNHSGE dataset and ChatGPT's performance on the AP Physics dataset from OpenAI~\cite{OpenAI_gpt_4_report}. OpenAI had reported ChatGPT's score range as $30 \%-66 \%$. The results showed that ChatGPT scored $61 \%$ on the VNHSGE dataset, while BingChat scored $66 \%$. The highest score of $78 \%$ was achieved by the test case, with a minimum of $30 \%$.

\subsubsection{Comparison to Vietnamese students}
 
To evaluate the performance of the LLMs, we compared their scores with those of Vietnamese students. The converted scores of ChatGPT and BingChat, as well as the average score (AVNS) and the most attained score by a Vietnamese student (MVNS), are shown in Table~\ref{tabl:Scores}. The average scores of ChatGPT and BingChat are similar to AVNS and lower scores than MVNS. However, Max performed better AVNS and MVNS. This shows the potential application of LLMs to high school physic in Vietnam.   

\begingroup
\renewcommand{\arraystretch}{1.5} 
\begin{table}[h!]
	\caption{Score}
	\label{tabl:Scores}
	\resizebox{0.5\textwidth}{!}{%
	\begin{tabular}{c|c|c|c|c|c|c|}
		\cline{2-7}
		\multicolumn{1}{l|}{}               & \textbf{ChatGPT} & \textbf{BingChat} & \textbf{Min} & \textbf{Max} & \textbf{AVNS} & \textbf{MVNS} \\ \hline
		\multicolumn{1}{|c|}{\textbf{2019}} & 6                & 5.5               & 4            & 7.5          & 5.57          & 6.25          \\ \hline
		\multicolumn{1}{|c|}{\textbf{2020}} & 6.25             & 6.75              & 5            & 8            & 6.72          & 7.75          \\ \hline
		\multicolumn{1}{|c|}{\textbf{2021}} & 6                & 6.75              & 4.75         & 8            & 6.56          & 7.5           \\ \hline
		\multicolumn{1}{|c|}{\textbf{2022}} & 6.5              & 6.75              & 5.5          & 7.75         & 6.72          & 7.25          \\ \hline
		\multicolumn{1}{|c|}{\textbf{2023}} & 5.75             & 7.25              & 5            & 8            &               &               \\ \hline
		\multicolumn{1}{|c|}{\textbf{AVG}}  & 6.1              & 6.6               & 4.85         & 7.85         & 6.39          &               \\ \hline
	\end{tabular}
	}
\end{table}
\endgroup

\begin{figure*}[ht!]
	\begin{center}
		\begin{tikzpicture}
			\begin{axis}[
				legend style={at={(0.5,-0.125)}, 	
					anchor=north,legend columns=-1}, 
				symbolic x coords={
					0,
					0.25,
					0.5,
					0.75,
					1,
					1.25,
					1.5,
					1.75,
					2,
					2.25,
					2.5,
					2.75,
					3,
					3.25,
					3.5,
					3.75,
					4,
					4.25,
					4.5,
					4.75,
					5,
					5.25,
					5.5,
					5.75,
					6,
					6.25,
					6.5,
					6.75,
					7,
					7.25,
					7.5,
					7.75,
					8,
					8.25,
					8.5,
					8.75,
					9,
					9.25,
					9.5,
					9.75,
					10,	
				},
				hide axis,
				ybar,
				bar width=5pt,
				ymin=0,
				ymax=20000,
				every node near coord/.append style={rotate=90, anchor=west},
				width=\textwidth, 
				enlarge x limits={abs=0.5*\pgfplotbarwidth},
				height=10cm, 
				width=16cm,
				axis x line*=bottom, axis y line*=left
				]
				\addplot [fill=green] coordinates {
					(0,0)
				};
				\addplot [fill=red] coordinates {
					(2.5,0)
				};	
				\addplot [fill=gray] coordinates {
					(5,0)
				};
				\addplot [fill=yellow] coordinates {
					(7.5,0)
				};	
				\addplot [fill=blue] coordinates {
					(10,0)
				};
				\legend{ChatGPT, BingChat, Min, Max, Vietnamese students}	
			\end{axis}
			
			\begin{axis}[
				symbolic x coords={
					0,
					0.25,
					0.5,
					0.75,
					1,
					1.25,
					1.5,
					1.75,
					2,
					2.25,
					2.5,
					2.75,
					3,
					3.25,
					3.5,
					3.75,
					4,
					4.25,
					4.5,
					4.75,
					5,
					5.25,
					5.5,
					5.75,
					6,
					6.25,
					6.5,
					6.75,
					7,
					7.25,
					7.5,
					7.75,
					8,
					8.25,
					8.5,
					8.75,
					9,
					9.25,
					9.5,
					9.75,
					10,	
				},
				hide axis,
				x tick label style={rotate=90,anchor=east},
				ybar,
				bar width=5pt,
				ymin=0,
				every node near coord/.append style={rotate=90, anchor=west},
				width=\textwidth, 
				height=9cm, 
				width=16cm,
				axis x line*=bottom, axis y line*=left
				]
				\addplot [fill=green] coordinates {
					(0,0)
					(0.25,0)
					(0.5,0)
					(0.75,0)
					(1,0)
					(1.25,0)
					(1.5,0)
					(1.75,0)
					(2,0)
					(2.25,0)
					(2.5,0)
					(2.75,0)
					(3,0)
					(3.25,0)
					(3.5,0)
					(3.75,0)
					(4,0)
					(4.25,0)
					(4.5,0)
					(4.75,0)
					(5,0)
					(5.25,0)
					(5.5,0)
					(5.75,0)
					(6,0)
					(6.25,30000)
					(6.5,0)
					(6.75,0)
					(7,0)
					(7.25,0)
					(7.5,0)
					(7.75,0)
					(8,0)
					(8.25,0)
					(8.5,0)
					(8.75,0)
					(9,0)
					(9.25,0)
					(9.5,0)
					(9.75,0)
					(10,0)
					
				};	
			\end{axis}
			
			\begin{axis}[ 
				symbolic x coords={
					0,
					0.25,
					0.5,
					0.75,
					1,
					1.25,
					1.5,
					1.75,
					2,
					2.25,
					2.5,
					2.75,
					3,
					3.25,
					3.5,
					3.75,
					4,
					4.25,
					4.5,
					4.75,
					5,
					5.25,
					5.5,
					5.75,
					6,
					6.25,
					6.5,
					6.75,
					7,
					7.25,
					7.5,
					7.75,
					8,
					8.25,
					8.5,
					8.75,
					9,
					9.25,
					9.5,
					9.75,
					10,	
				},
				hide axis,
				ybar,
				bar width=5pt,
				ymin=0,
				every node near coord/.append style={rotate=90, anchor=west},
				width=\textwidth, 
				height=9cm, 
				width=16cm,
				axis x line*=bottom, axis y line*=left
				]
				\addplot [fill=red] coordinates {
					(0,0)
					(0.25,0)
					(0.5,0)
					(0.75,0)
					(1,0)
					(1.25,0)
					(1.5,0)
					(1.75,0)
					(2,0)
					(2.25,0)
					(2.5,0)
					(2.75,0)
					(3,0)
					(3.25,0)
					(3.5,0)
					(3.75,0)
					(4,0)
					(4.25,0)
					(4.5,0)
					(4.75,0)
					(5,0)
					(5.25,0)
					(5.5,0)
					(5.75,0)
					(6,0)
					(6.25,0)
					(6.5,0)
					(6.75,30000)
					(7,0)
					(7.25,0)
					(7.5,0)
					(7.75,0)
					(8,0)
					(8.25,0)
					(8.5,0)
					(8.75,0)
					(9,0)
					(9.25,0)
					(9.5,0)
					(9.75,0)
					(10,0)
					
				};	
			\end{axis}
			
			\begin{axis}[ 
				symbolic x coords={
					0,
					0.25,
					0.5,
					0.75,
					1,
					1.25,
					1.5,
					1.75,
					2,
					2.25,
					2.5,
					2.75,
					3,
					3.25,
					3.5,
					3.75,
					4,
					4.25,
					4.5,
					4.75,
					5,
					5.25,
					5.5,
					5.75,
					6,
					6.25,
					6.5,
					6.75,
					7,
					7.25,
					7.5,
					7.75,
					8,
					8.25,
					8.5,
					8.75,
					9,
					9.25,
					9.5,
					9.75,
					10,	
				},
				hide axis,
				ybar,
				bar width=5pt,
				ymin=0,
				every node near coord/.append style={rotate=90, anchor=west},
				width=\textwidth, 
				height=9cm, 
				width=16cm,
				axis x line*=bottom, axis y line*=left
				]
				\addplot [fill=gray] coordinates {
					(0,0)
					(0.25,0)
					(0.5,0)
					(0.75,0)
					(1,0)
					(1.25,0)
					(1.5,0)
					(1.75,0)
					(2,0)
					(2.25,0)
					(2.5,0)
					(2.75,0)
					(3,0)
					(3.25,0)
					(3.5,0)
					(3.75,0)
					(4,0)
					(4.25,0)
					(4.5,0)
					(4.75,0)
					(5,30000)
					(5.25,0)
					(5.5,0)
					(5.75,0)
					(6,0)
					(6.25,0)
					(6.5,0)
					(6.75,0)
					(7,0)
					(7.25,0)
					(7.5,0)
					(7.75,0)
					(8,0)
					(8.25,0)
					(8.5,0)
					(8.75,0)
					(9,0)
					(9.25,0)
					(9.5,0)
					(9.75,0)
					(10,0)	
				};	
			\end{axis}
			
			\begin{axis}[ 
				symbolic x coords={
					0,
					0.25,
					0.5,
					0.75,
					1,
					1.25,
					1.5,
					1.75,
					2,
					2.25,
					2.5,
					2.75,
					3,
					3.25,
					3.5,
					3.75,
					4,
					4.25,
					4.5,
					4.75,
					5,
					5.25,
					5.5,
					5.75,
					6,
					6.25,
					6.5,
					6.75,
					7,
					7.25,
					7.5,
					7.75,
					8,
					8.25,
					8.5,
					8.75,
					9,
					9.25,
					9.5,
					9.75,
					10,	
				},
				hide axis,
				ybar,
				bar width=5pt,
				ymin=0,
				every node near coord/.append style={rotate=90, anchor=west},
				width=\textwidth, 
				height=9cm, 
				width=16cm,
				axis x line*=bottom, axis y line*=left
				]
				\addplot [fill=yellow] coordinates {
					(0,0)
					(0.25,0)
					(0.5,0)
					(0.75,0)
					(1,0)
					(1.25,0)
					(1.5,0)
					(1.75,0)
					(2,0)
					(2.25,0)
					(2.5,0)
					(2.75,0)
					(3,0)
					(3.25,0)
					(3.5,0)
					(3.75,0)
					(4,0)
					(4.25,0)
					(4.5,0)
					(4.75,0)
					(5,0)
					(5.25,0)
					(5.5,0)
					(5.75,0)
					(6,0)
					(6.25,0)
					(6.5,0)
					(6.75,0)
					(7,0)
					(7.25,0)
					(7.5,0)
					(7.75,0)
					(8,30000)
					(8.25,0)
					(8.5,0)
					(8.75,0)
					(9,0)
					(9.25,0)
					(9.5,0)
					(9.75,0)
					(10,0)
					
				};	
			\end{axis}

			\begin{axis}[
				ylabel={Number of Student},
				symbolic x coords={
					0,
					0.25,
					0.5,
					0.75,
					1,
					1.25,
					1.5,
					1.75,
					2,
					2.25,
					2.5,
					2.75,
					3,
					3.25,
					3.5,
					3.75,
					4,
					4.25,
					4.5,
					4.75,
					5,
					5.25,
					5.5,
					5.75,
					6,
					6.25,
					6.5,
					6.75,
					7,
					7.25,
					7.5,
					7.75,
					8,
					8.25,
					8.5,
					8.75,
					9,
					9.25,
					9.5,
					9.75,
					10,	
				},
				xtick=data,
				x tick label style={rotate=90,anchor=east},
				ybar,
				bar width=5pt,
				ymin=0,
				nodes near coords,   
				every node near coord/.append style={rotate=90, anchor=west},
				width=\textwidth, 
				height=9cm, 
				width=16cm,
				axis x line*=bottom, axis y line*=left
				]
				\addplot [fill=blue] coordinates {
					(0,8)
					(0.25,0)
					(0.5,1)
					(0.75,10)
					(1,20)
					(1.25,69)
					(1.5,170)
					(1.75,292)
					(2,485)
					(2.25,801)
					(2.5,1146)
					(2.75,1560)
					(3,2099)
					(3.25,2555)
					(3.5,3100)
					(3.75,3570)
					(4,4321)
					(4.25,4903)
					(4.5,5627)
					(4.75,6403)
					(5,7350)
					(5.25,8387)
					(5.5,9498)
					(5.75,10706)
					(6,12252)
					(6.25,13590)
					(6.5,15260)
					(6.75,17063)
					(7,19295)
					(7.25,21386)
					(7.5,22986)
					(7.75,23214)
					(8,21588)
					(8.25,18290)
					(8.5,13406)
					(8.75,8314)
					(9,4588)
					(9.25,1847)
					(9.5,541)
					(9.75,136)
					(10,10)
					
				};	
				
			\end{axis}
		\end{tikzpicture}
	\end{center}
	\caption{Comparison between ChatGPT-BingChat performances and Vietnamese students in 2020.}
	\label{fig:phy_2020}
\end{figure*}
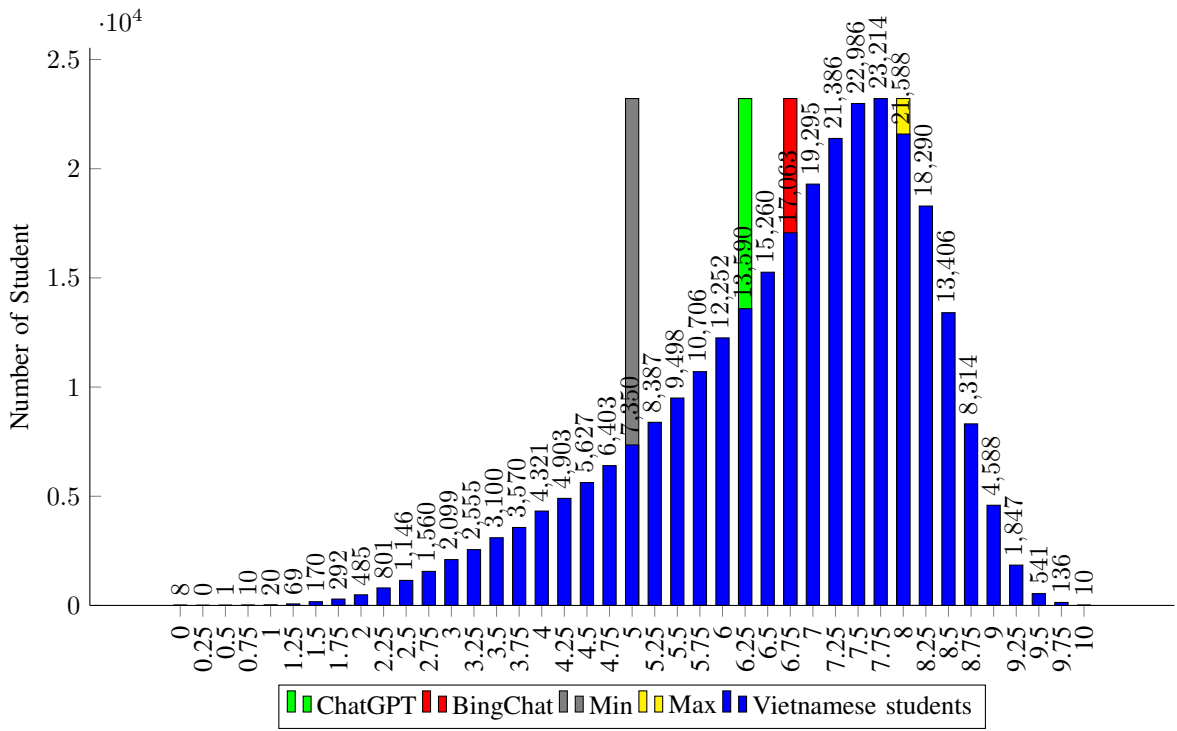

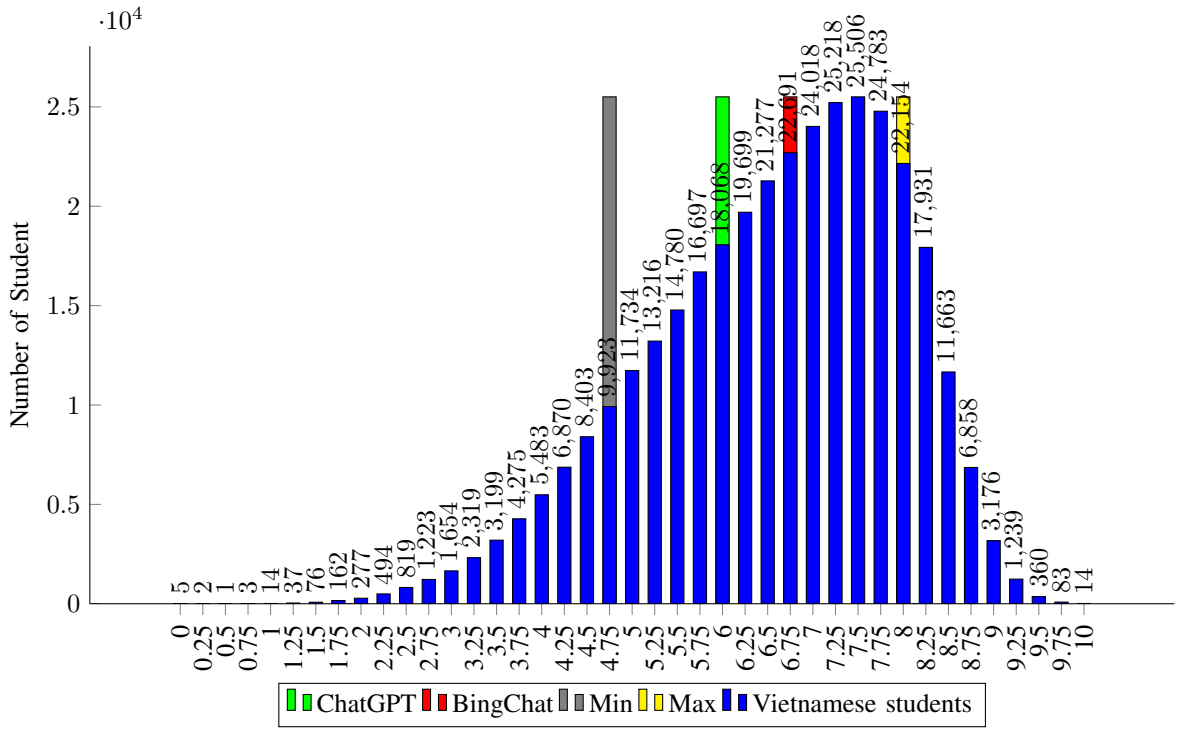
\begin{figure*}[ht!]
	\begin{center}
		\begin{tikzpicture}
			\begin{axis}[
				legend style={at={(0.5,-0.125)}, 	
					anchor=north,legend columns=-1}, 
				symbolic x coords={
					0,
					0.25,
					0.5,
					0.75,
					1,
					1.25,
					1.5,
					1.75,
					2,
					2.25,
					2.5,
					2.75,
					3,
					3.25,
					3.5,
					3.75,
					4,
					4.25,
					4.5,
					4.75,
					5,
					5.25,
					5.5,
					5.75,
					6,
					6.25,
					6.5,
					6.75,
					7,
					7.25,
					7.5,
					7.75,
					8,
					8.25,
					8.5,
					8.75,
					9,
					9.25,
					9.5,
					9.75,
					10,	
				},
				hide axis,
				ybar,
				bar width=5pt,
				ymin=0,
				ymax=1000,
				every node near coord/.append style={rotate=90, anchor=west},
				width=\textwidth, 
				enlarge x limits={abs=0.5*\pgfplotbarwidth},
				height=10cm, 
				width=16cm,
				axis x line*=bottom, axis y line*=left
				]
				\addplot [fill=green] coordinates {
					(0,0)
				};
				\addplot [fill=red] coordinates {
					(2.5,0)
				};	
				\addplot [fill=gray] coordinates {
					(5,0)
				};
				\addplot [fill=yellow] coordinates {
					(7.5,0)
				};	
				\addplot [fill=blue] coordinates {
					(10,0)
				};
				\legend{ChatGPT, BingChat, Min, Max, Vietnamese students}
			\end{axis}
			
			\begin{axis}[
				symbolic x coords={
					0,
					0.25,
					0.5,
					0.75,
					1,
					1.25,
					1.5,
					1.75,
					2,
					2.25,
					2.5,
					2.75,
					3,
					3.25,
					3.5,
					3.75,
					4,
					4.25,
					4.5,
					4.75,
					5,
					5.25,
					5.5,
					5.75,
					6,
					6.25,
					6.5,
					6.75,
					7,
					7.25,
					7.5,
					7.75,
					8,
					8.25,
					8.5,
					8.75,
					9,
					9.25,
					9.5,
					9.75,
					10,	
				},
				hide axis,
				x tick label style={rotate=90,anchor=east},
				ybar,
				bar width=5pt,
				ymin=0,
				every node near coord/.append style={rotate=90, anchor=west},
				width=\textwidth, 
				height=9cm, 
				width=16cm,
				axis x line*=bottom, axis y line*=left
				]
				\addplot [fill=green] coordinates {
					(0,0)
					(0.25,0)
					(0.5,0)
					(0.75,0)
					(1,0)
					(1.25,0)
					(1.5,0)
					(1.75,0)
					(2,0)
					(2.25,0)
					(2.5,0)
					(2.75,0)
					(3,0)
					(3.25,0)
					(3.5,0)
					(3.75,0)
					(4,0)
					(4.25,0)
					(4.5,0)
					(4.75,0)
					(5,0)
					(5.25,0)
					(5.5,0)
					(5.75,0)
					(6,30000)
					(6.25,0)
					(6.5,0)
					(6.75,0)
					(7,0)
					(7.25,0)
					(7.5,0)
					(7.75,0)
					(8,0)
					(8.25,0)
					(8.5,0)
					(8.75,0)
					(9,0)
					(9.25,0)
					(9.5,0)
					(9.75,0)
					(10,0)
					
				};	
			\end{axis}
			
			\begin{axis}[ 
				symbolic x coords={
					0,
					0.25,
					0.5,
					0.75,
					1,
					1.25,
					1.5,
					1.75,
					2,
					2.25,
					2.5,
					2.75,
					3,
					3.25,
					3.5,
					3.75,
					4,
					4.25,
					4.5,
					4.75,
					5,
					5.25,
					5.5,
					5.75,
					6,
					6.25,
					6.5,
					6.75,
					7,
					7.25,
					7.5,
					7.75,
					8,
					8.25,
					8.5,
					8.75,
					9,
					9.25,
					9.5,
					9.75,
					10,	
				},
				hide axis,
				ybar,
				bar width=5pt,
				ymin=0,
				every node near coord/.append style={rotate=90, anchor=west},
				width=\textwidth, 
				height=9cm, 
				width=16cm,
				axis x line*=bottom, axis y line*=left
				]
				\addplot [fill=red] coordinates {
					(0,0)
					(0.25,0)
					(0.5,0)
					(0.75,0)
					(1,0)
					(1.25,0)
					(1.5,0)
					(1.75,0)
					(2,0)
					(2.25,0)
					(2.5,0)
					(2.75,0)
					(3,0)
					(3.25,0)
					(3.5,0)
					(3.75,0)
					(4,0)
					(4.25,0)
					(4.5,0)
					(4.75,0)
					(5,0)
					(5.25,0)
					(5.5,0)
					(5.75,0)
					(6,0)
					(6.25,0)
					(6.5,0)
					(6.75,30000)
					(7,0)
					(7.25,0)
					(7.5,0)
					(7.75,0)
					(8,0)
					(8.25,0)
					(8.5,0)
					(8.75,0)
					(9,0)
					(9.25,0)
					(9.5,0)
					(9.75,0)
					(10,0)
					
				};	
			\end{axis}
			
			\begin{axis}[ 
				symbolic x coords={
					0,
					0.25,
					0.5,
					0.75,
					1,
					1.25,
					1.5,
					1.75,
					2,
					2.25,
					2.5,
					2.75,
					3,
					3.25,
					3.5,
					3.75,
					4,
					4.25,
					4.5,
					4.75,
					5,
					5.25,
					5.5,
					5.75,
					6,
					6.25,
					6.5,
					6.75,
					7,
					7.25,
					7.5,
					7.75,
					8,
					8.25,
					8.5,
					8.75,
					9,
					9.25,
					9.5,
					9.75,
					10,	
				},
				hide axis,
				ybar,
				bar width=5pt,
				ymin=0,
				every node near coord/.append style={rotate=90, anchor=west},
				width=\textwidth, 
				height=9cm, 
				width=16cm,
				axis x line*=bottom, axis y line*=left
				]
				\addplot [fill=gray] coordinates {
					(0,0)
					(0.25,0)
					(0.5,0)
					(0.75,0)
					(1,0)
					(1.25,0)
					(1.5,0)
					(1.75,0)
					(2,0)
					(2.25,0)
					(2.5,0)
					(2.75,0)
					(3,0)
					(3.25,0)
					(3.5,0)
					(3.75,0)
					(4,0)
					(4.25,0)
					(4.5,0)
					(4.75,30000)
					(5,0)
					(5.25,0)
					(5.5,0)
					(5.75,0)
					(6,0)
					(6.25,0)
					(6.5,0)
					(6.75,0)
					(7,0)
					(7.25,0)
					(7.5,0)
					(7.75,0)
					(8,0)
					(8.25,0)
					(8.5,0)
					(8.75,0)
					(9,0)
					(9.25,0)
					(9.5,0)
					(9.75,0)
					(10,0)
					
				};	
			\end{axis}
			
			\begin{axis}[ 
				symbolic x coords={
					0,
					0.25,
					0.5,
					0.75,
					1,
					1.25,
					1.5,
					1.75,
					2,
					2.25,
					2.5,
					2.75,
					3,
					3.25,
					3.5,
					3.75,
					4,
					4.25,
					4.5,
					4.75,
					5,
					5.25,
					5.5,
					5.75,
					6,
					6.25,
					6.5,
					6.75,
					7,
					7.25,
					7.5,
					7.75,
					8,
					8.25,
					8.5,
					8.75,
					9,
					9.25,
					9.5,
					9.75,
					10,	
				},
				hide axis,
				ybar,
				bar width=5pt,
				ymin=0,
				every node near coord/.append style={rotate=90, anchor=west},
				width=\textwidth, 
				height=9cm, 
				width=16cm,
				axis x line*=bottom, axis y line*=left
				]
				\addplot [fill=yellow] coordinates {
					(0,0)
					(0.25,0)
					(0.5,0)
					(0.75,0)
					(1,0)
					(1.25,0)
					(1.5,0)
					(1.75,0)
					(2,0)
					(2.25,0)
					(2.5,0)
					(2.75,0)
					(3,0)
					(3.25,0)
					(3.5,0)
					(3.75,0)
					(4,0)
					(4.25,0)
					(4.5,0)
					(4.75,0)
					(5,0)
					(5.25,0)
					(5.5,0)
					(5.75,0)
					(6,0)
					(6.25,0)
					(6.5,0)
					(6.75,0)
					(7,0)
					(7.25,0)
					(7.5,0)
					(7.75,0)
					(8,30000)
					(8.25,0)
					(8.5,0)
					(8.75,0)
					(9,0)
					(9.25,0)
					(9.5,0)
					(9.75,0)
					(10,0)
					
				};	
			\end{axis}
			
			\begin{axis}[
				ylabel={Number of Student},
				symbolic x coords={
					0,
					0.25,
					0.5,
					0.75,
					1,
					1.25,
					1.5,
					1.75,
					2,
					2.25,
					2.5,
					2.75,
					3,
					3.25,
					3.5,
					3.75,
					4,
					4.25,
					4.5,
					4.75,
					5,
					5.25,
					5.5,
					5.75,
					6,
					6.25,
					6.5,
					6.75,
					7,
					7.25,
					7.5,
					7.75,
					8,
					8.25,
					8.5,
					8.75,
					9,
					9.25,
					9.5,
					9.75,
					10,	
				},
				xtick=data,
				x tick label style={rotate=90,anchor=east},
				ybar,
				bar width=5pt,
				ymin=0,
				nodes near coords,   
				every node near coord/.append style={rotate=90, anchor=west},
				width=\textwidth, 
				height=9cm, 
				width=16cm,
				axis x line*=bottom, axis y line*=left
				]
				\addplot [fill=blue] coordinates {
					(0,5)
					(0.25,2)
					(0.5,1)
					(0.75,3)
					(1,14)
					(1.25,37)
					(1.5,76)
					(1.75,162)
					(2,277)
					(2.25,494)
					(2.5,819)
					(2.75,1223)
					(3,1654)
					(3.25,2319)
					(3.5,3199)
					(3.75,4275)
					(4,5483)
					(4.25,6870)
					(4.5,8403)
					(4.75,9923)
					(5,11734)
					(5.25,13216)
					(5.5,14780)
					(5.75,16697)
					(6,18068)
					(6.25,19699)
					(6.5,21277)
					(6.75,22691)
					(7,24018)
					(7.25,25218)
					(7.5,25506)
					(7.75,24783)
					(8,22154)
					(8.25,17931)
					(8.5,11663)
					(8.75,6858)
					(9,3176)
					(9.25,1239)
					(9.5,360)
					(9.75,83)
					(10,14)

				};	
				
			\end{axis}
		\end{tikzpicture}
	\end{center}
	\caption{Comparison between ChatGPT-BingChat performances and Vietnamese students in 2021.}
	\label{fig:phy_2021}
\end{figure*}

\begin{figure*}[ht!]
	\begin{center}
		\begin{tikzpicture}
			\begin{axis}[
				legend style={at={(0.5,-0.125)}, 	
					anchor=north,legend columns=-1}, 
				symbolic x coords={
					0,
					0.25,
					0.5,
					0.75,
					1,
					1.25,
					1.5,
					1.75,
					2,
					2.25,
					2.5,
					2.75,
					3,
					3.25,
					3.5,
					3.75,
					4,
					4.25,
					4.5,
					4.75,
					5,
					5.25,
					5.5,
					5.75,
					6,
					6.25,
					6.5,
					6.75,
					7,
					7.25,
					7.5,
					7.75,
					8,
					8.25,
					8.5,
					8.75,
					9,
					9.25,
					9.5,
					9.75,
					10,	
				},
				hide axis,
				ybar,
				bar width=5pt,
				ymin=0,
				ymax=1000,
				every node near coord/.append style={rotate=90, anchor=west},
				width=\textwidth, 
				enlarge x limits={abs=0.5*\pgfplotbarwidth},
				height=10cm, 
				width=16cm,
				axis x line*=bottom, axis y line*=left
				]
				\addplot [fill=green] coordinates {
					(0,0)
				};
				\addplot [fill=red] coordinates {
					(2.5,0)
				};	
				\addplot [fill=gray] coordinates {
					(5,0)
				};
				\addplot [fill=yellow] coordinates {
					(7.5,0)
				};	
				\addplot [fill=blue] coordinates {
					(10,0)
				};
				\legend{ChatGPT, BingChat, Min, Max, Vietnamese students}	
			\end{axis}
			
			\begin{axis}[
				symbolic x coords={
					0,
					0.25,
					0.5,
					0.75,
					1,
					1.25,
					1.5,
					1.75,
					2,
					2.25,
					2.5,
					2.75,
					3,
					3.25,
					3.5,
					3.75,
					4,
					4.25,
					4.5,
					4.75,
					5,
					5.25,
					5.5,
					5.75,
					6,
					6.25,
					6.5,
					6.75,
					7,
					7.25,
					7.5,
					7.75,
					8,
					8.25,
					8.5,
					8.75,
					9,
					9.25,
					9.5,
					9.75,
					10,	
				},
				hide axis,
				x tick label style={rotate=90,anchor=east},
				ybar,
				bar width=5pt,
				ymin=0,
				every node near coord/.append style={rotate=90, anchor=west},
				width=\textwidth, 
				height=9cm, 
				width=16cm,
				axis x line*=bottom, axis y line*=left
				]
				\addplot [fill=green] coordinates {
					(0,0)
					(0.25,0)
					(0.5,0)
					(0.75,0)
					(1,0)
					(1.25,0)
					(1.5,0)
					(1.75,0)
					(2,0)
					(2.25,0)
					(2.5,0)
					(2.75,0)
					(3,0)
					(3.25,0)
					(3.5,0)
					(3.75,0)
					(4,0)
					(4.25,0)
					(4.5,0)
					(4.75,0)
					(5,0)
					(5.25,0)
					(5.5,0)
					(5.75,0)
					(6,0)
					(6.25,0)
					(6.5,30000)
					(6.75,0)
					(7,0)
					(7.25,0)
					(7.5,0)
					(7.75,0)
					(8,0)
					(8.25,0)
					(8.5,0)
					(8.75,0)
					(9,0)
					(9.25,0)
					(9.5,0)
					(9.75,0)
					(10,0)
					
				};	
			\end{axis}
			
			\begin{axis}[ 
				symbolic x coords={
					0,
					0.25,
					0.5,
					0.75,
					1,
					1.25,
					1.5,
					1.75,
					2,
					2.25,
					2.5,
					2.75,
					3,
					3.25,
					3.5,
					3.75,
					4,
					4.25,
					4.5,
					4.75,
					5,
					5.25,
					5.5,
					5.75,
					6,
					6.25,
					6.5,
					6.75,
					7,
					7.25,
					7.5,
					7.75,
					8,
					8.25,
					8.5,
					8.75,
					9,
					9.25,
					9.5,
					9.75,
					10,	
				},
				hide axis,
				ybar,
				bar width=5pt,
				ymin=0,
				every node near coord/.append style={rotate=90, anchor=west},
				width=\textwidth, 
				height=9cm, 
				width=16cm,
				axis x line*=bottom, axis y line*=left
				]
				\addplot [fill=red] coordinates {
					(0,0)
					(0.25,0)
					(0.5,0)
					(0.75,0)
					(1,0)
					(1.25,0)
					(1.5,0)
					(1.75,0)
					(2,0)
					(2.25,0)
					(2.5,0)
					(2.75,0)
					(3,0)
					(3.25,0)
					(3.5,0)
					(3.75,0)
					(4,0)
					(4.25,0)
					(4.5,0)
					(4.75,0)
					(5,0)
					(5.25,0)
					(5.5,0)
					(5.75,0)
					(6,0)
					(6.25,0)
					(6.5,0)
					(6.75,30000)
					(7,0)
					(7.25,0)
					(7.5,0)
					(7.75,0)
					(8,0)
					(8.25,0)
					(8.5,0)
					(8.75,0)
					(9,0)
					(9.25,0)
					(9.5,0)
					(9.75,0)
					(10,0)
					
				};	
			\end{axis}
			
			\begin{axis}[ 
				symbolic x coords={
					0,
					0.25,
					0.5,
					0.75,
					1,
					1.25,
					1.5,
					1.75,
					2,
					2.25,
					2.5,
					2.75,
					3,
					3.25,
					3.5,
					3.75,
					4,
					4.25,
					4.5,
					4.75,
					5,
					5.25,
					5.5,
					5.75,
					6,
					6.25,
					6.5,
					6.75,
					7,
					7.25,
					7.5,
					7.75,
					8,
					8.25,
					8.5,
					8.75,
					9,
					9.25,
					9.5,
					9.75,
					10,	
				},
				hide axis,
				ybar,
				bar width=5pt,
				ymin=0,
				every node near coord/.append style={rotate=90, anchor=west},
				width=\textwidth, 
				height=9cm, 
				width=16cm,
				axis x line*=bottom, axis y line*=left
				]
				\addplot [fill=gray] coordinates {
					(0,0)
					(0.25,0)
					(0.5,0)
					(0.75,0)
					(1,0)
					(1.25,0)
					(1.5,0)
					(1.75,0)
					(2,0)
					(2.25,0)
					(2.5,0)
					(2.75,0)
					(3,0)
					(3.25,0)
					(3.5,0)
					(3.75,0)
					(4,0)
					(4.25,0)
					(4.5,0)
					(4.75,0)
					(5,0)
					(5.25,0)
					(5.5,30000)
					(5.75,0)
					(6,0)
					(6.25,0)
					(6.5,0)
					(6.75,0)
					(7,0)
					(7.25,0)
					(7.5,0)
					(7.75,0)
					(8,0)
					(8.25,0)
					(8.5,0)
					(8.75,0)
					(9,0)
					(9.25,0)
					(9.5,0)
					(9.75,0)
					(10,0)
					
				};	
			\end{axis}
			
			\begin{axis}[ 
				symbolic x coords={
					0,
					0.25,
					0.5,
					0.75,
					1,
					1.25,
					1.5,
					1.75,
					2,
					2.25,
					2.5,
					2.75,
					3,
					3.25,
					3.5,
					3.75,
					4,
					4.25,
					4.5,
					4.75,
					5,
					5.25,
					5.5,
					5.75,
					6,
					6.25,
					6.5,
					6.75,
					7,
					7.25,
					7.5,
					7.75,
					8,
					8.25,
					8.5,
					8.75,
					9,
					9.25,
					9.5,
					9.75,
					10,	
				},
				hide axis,
				ybar,
				bar width=5pt,
				ymin=0,
				every node near coord/.append style={rotate=90, anchor=west},
				width=\textwidth, 
				height=9cm, 
				width=16cm,
				axis x line*=bottom, axis y line*=left
				]
				\addplot [fill=yellow] coordinates {
					(0,0)
					(0.25,0)
					(0.5,0)
					(0.75,0)
					(1,0)
					(1.25,0)
					(1.5,0)
					(1.75,0)
					(2,0)
					(2.25,0)
					(2.5,0)
					(2.75,0)
					(3,0)
					(3.25,0)
					(3.5,0)
					(3.75,0)
					(4,0)
					(4.25,0)
					(4.5,0)
					(4.75,0)
					(5,0)
					(5.25,0)
					(5.5,0)
					(5.75,0)
					(6,0)
					(6.25,0)
					(6.5,0)
					(6.75,0)
					(7,0)
					(7.25,0)
					(7.5,30000)
					(7.75,0)
					(8,0)
					(8.25,0)
					(8.5,0)
					(8.75,0)
					(9,0)
					(9.25,0)
					(9.5,0)
					(9.75,0)
					(10,0)
					
				};	
			\end{axis}

			\begin{axis}[
				ylabel={Number of Student},
				symbolic x coords={
					0,
					0.25,
					0.5,
					0.75,
					1,
					1.25,
					1.5,
					1.75,
					2,
					2.25,
					2.5,
					2.75,
					3,
					3.25,
					3.5,
					3.75,
					4,
					4.25,
					4.5,
					4.75,
					5,
					5.25,
					5.5,
					5.75,
					6,
					6.25,
					6.5,
					6.75,
					7,
					7.25,
					7.5,
					7.75,
					8,
					8.25,
					8.5,
					8.75,
					9,
					9.25,
					9.5,
					9.75,
					10,	
				},
				xtick=data,
				x tick label style={rotate=90,anchor=east},
				ybar,
				bar width=5pt,
				ymin=0,
				nodes near coords,   
				every node near coord/.append style={rotate=90, anchor=west},
				width=\textwidth, 
				height=9cm, 
				width=16cm,
				axis x line*=bottom, axis y line*=left
				]
				\addplot [fill=blue] coordinates {
					(0,12)
					(0.25,0)
					(0.5,0)
					(0.75,1)
					(1,11)
					(1.25,24)
					(1.5,71)
					(1.75,138)
					(2,241)
					(2.25,388)
					(2.5,645)
					(2.75,1029)
					(3,1526)
					(3.25,2031)
					(3.5,2818)
					(3.75,3581)
					(4,4613)
					(4.25,5661)
					(4.5,7005)
					(4.75,8252)
					(5,9549)
					(5.25,11160)
					(5.5,12696)
					(5.75,14556)
					(6,16355)
					(6.25,17962)
					(6.5,19475)
					(6.75,21240)
					(7,22546)
					(7.25,23162)
					(7.5,22883)
					(7.75,21849)
					(8,19664)
					(8.25,16373)
					(8.5,13432)
					(8.75,10232)
					(9,7192)
					(9.25,4278)
					(9.5,2012)
					(9.75,708)
					(10,154)
					
				};	
				
			\end{axis}
		\end{tikzpicture}
	\end{center}
	\caption{Comparison between ChatGPT-BingChat performances and Vietnamese students in 2022.}
	\label{fig:phy_2022}
\end{figure*}
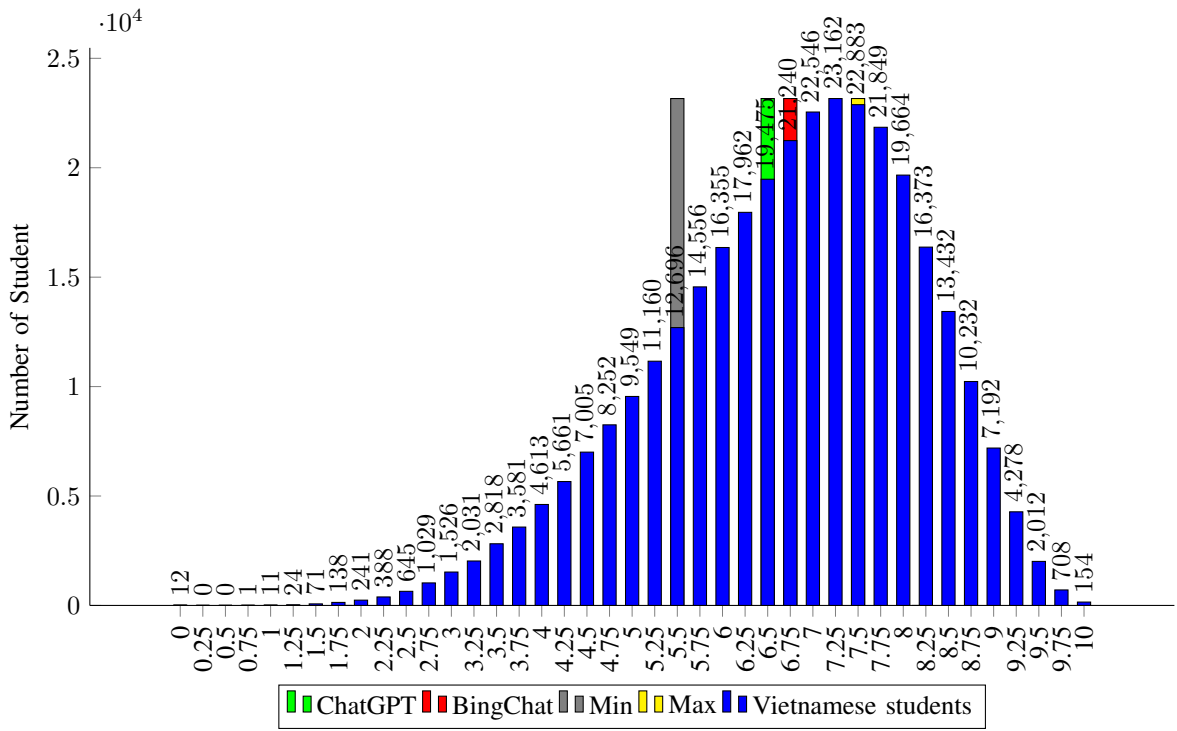

In our study, we evaluated the performance of ChatGPT, BingChat, Min, and Max on high school physics exams and compared their scores with those of Vietnamese students. The results indicated that although LLMs have made significant strides in NLP, their performance in specialized domains like physics still falls short of that of human students. Fig~\ref{fig:phy_2019}-\ref{fig:phy_2022} show the physics score spectrum of Vietnamese students in 2019-2022. ChatGPT and BingGPT's performance was mostly inferior to that of Vietnamese students. This underscores the need to further refine and optimize these models for specialized domains to achieve human-level performance. Additionally, future research could investigate ways to incorporate domain-specific knowledge and curriculum into the training of these models to improve their performance on subject-specific exams.

\begin{figure*}[ht!]
	\begin{center}
		\begin{tikzpicture}
			\begin{axis}[
				legend style={at={(0.5,-0.125)}, 	
					anchor=north,legend columns=-1}, 
				symbolic x coords={
					0,
					0.25,
					0.5,
					0.75,
					1,
					1.25,
					1.5,
					1.75,
					2,
					2.25,
					2.5,
					2.75,
					3,
					3.25,
					3.5,
					3.75,
					4,
					4.25,
					4.5,
					4.75,
					5,
					5.25,
					5.5,
					5.75,
					6,
					6.25,
					6.5,
					6.75,
					7,
					7.25,
					7.5,
					7.75,
					8,
					8.25,
					8.5,
					8.75,
					9,
					9.25,
					9.5,
					9.75,
					10,	
				},
				hide axis,
				ybar,
				bar width=5pt,
				ymin=0,
				ymax=1000,
				every node near coord/.append style={rotate=90, anchor=west},
				width=\textwidth, 
				enlarge x limits={abs=0.5*\pgfplotbarwidth},
				height=10cm, 
				width=16cm,
				axis x line*=bottom, axis y line*=left
				]
				\addplot [fill=green] coordinates {
					(0,0)
				};
				\addplot [fill=red] coordinates {
					(2.5,0)
				};	
				\addplot [fill=gray] coordinates {
					(5,0)
				};
				\addplot [fill=yellow] coordinates {
					(7.5,0)
				};	
				\addplot [fill=blue] coordinates {
					(10,0)
				};
				\legend{ChatGPT, BingChat, Min, Max, Vietnamese students}
			\end{axis}
			
			\begin{axis}[
				symbolic x coords={
					0,
					0.25,
					0.5,
					0.75,
					1,
					1.25,
					1.5,
					1.75,
					2,
					2.25,
					2.5,
					2.75,
					3,
					3.25,
					3.5,
					3.75,
					4,
					4.25,
					4.5,
					4.75,
					5,
					5.25,
					5.5,
					5.75,
					6,
					6.25,
					6.5,
					6.75,
					7,
					7.25,
					7.5,
					7.75,
					8,
					8.25,
					8.5,
					8.75,
					9,
					9.25,
					9.5,
					9.75,
					10,	
				},
				hide axis,
				x tick label style={rotate=90,anchor=east},
				ybar,
				bar width=5pt,
				ymin=0,
				every node near coord/.append style={rotate=90, anchor=west},
				width=\textwidth, 
				height=9cm, 
				width=16cm,
				axis x line*=bottom, axis y line*=left
				]
				\addplot [fill=green] coordinates {
					(0,0)
					(0.25,0)
					(0.5,0)
					(0.75,0)
					(1,0)
					(1.25,0)
					(1.5,0)
					(1.75,0)
					(2,0)
					(2.25,0)
					(2.5,0)
					(2.75,0)
					(3,0)
					(3.25,0)
					(3.5,0)
					(3.75,0)
					(4,0)
					(4.25,0)
					(4.5,0)
					(4.75,0)
					(5,0)
					(5.25,0)
					(5.5,0)
					(5.75,0)
					(6,30000)
					(6.25,0)
					(6.5,0)
					(6.75,0)
					(7,0)
					(7.25,0)
					(7.5,0)
					(7.75,0)
					(8,0)
					(8.25,0)
					(8.5,0)
					(8.75,0)
					(9,0)
					(9.25,0)
					(9.5,0)
					(9.75,0)
					(10,0)
					
				};	
			\end{axis}
			
			\begin{axis}[ 
				symbolic x coords={
					0,
					0.25,
					0.5,
					0.75,
					1,
					1.25,
					1.5,
					1.75,
					2,
					2.25,
					2.5,
					2.75,
					3,
					3.25,
					3.5,
					3.75,
					4,
					4.25,
					4.5,
					4.75,
					5,
					5.25,
					5.5,
					5.75,
					6,
					6.25,
					6.5,
					6.75,
					7,
					7.25,
					7.5,
					7.75,
					8,
					8.25,
					8.5,
					8.75,
					9,
					9.25,
					9.5,
					9.75,
					10,	
				},
				hide axis,
				ybar,
				bar width=5pt,
				ymin=0,
				every node near coord/.append style={rotate=90, anchor=west},
				width=\textwidth, 
				height=9cm, 
				width=16cm,
				axis x line*=bottom, axis y line*=left
				]
				\addplot [fill=red] coordinates {
					(0,0)
					(0.25,0)
					(0.5,0)
					(0.75,0)
					(1,0)
					(1.25,0)
					(1.5,0)
					(1.75,0)
					(2,0)
					(2.25,0)
					(2.5,0)
					(2.75,0)
					(3,0)
					(3.25,0)
					(3.5,0)
					(3.75,0)
					(4,0)
					(4.25,0)
					(4.5,0)
					(4.75,0)
					(5,0)
					(5.25,0)
					(5.5,30000)
					(5.75,0)
					(6,0)
					(6.25,0)
					(6.5,0)
					(6.75,0)
					(7,0)
					(7.25,0)
					(7.5,0)
					(7.75,0)
					(8,0)
					(8.25,0)
					(8.5,0)
					(8.75,0)
					(9,0)
					(9.25,0)
					(9.5,0)
					(9.75,0)
					(10,0)
					
				};	
			\end{axis}
			
			\begin{axis}[ 
				symbolic x coords={
					0,
					0.25,
					0.5,
					0.75,
					1,
					1.25,
					1.5,
					1.75,
					2,
					2.25,
					2.5,
					2.75,
					3,
					3.25,
					3.5,
					3.75,
					4,
					4.25,
					4.5,
					4.75,
					5,
					5.25,
					5.5,
					5.75,
					6,
					6.25,
					6.5,
					6.75,
					7,
					7.25,
					7.5,
					7.75,
					8,
					8.25,
					8.5,
					8.75,
					9,
					9.25,
					9.5,
					9.75,
					10,	
				},
				hide axis,
				ybar,
				bar width=5pt,
				ymin=0,
				every node near coord/.append style={rotate=90, anchor=west},
				width=\textwidth, 
				height=9cm, 
				width=16cm,
				axis x line*=bottom, axis y line*=left
				]
				\addplot [fill=gray] coordinates {
					(0,0)
					(0.25,0)
					(0.5,0)
					(0.75,0)
					(1,0)
					(1.25,0)
					(1.5,0)
					(1.75,0)
					(2,0)
					(2.25,0)
					(2.5,0)
					(2.75,0)
					(3,0)
					(3.25,0)
					(3.5,0)
					(3.75,0)
					(4,30000)
					(4.25,0)
					(4.5,0)
					(4.75,0)
					(5,0)
					(5.25,0)
					(5.5,0)
					(5.75,0)
					(6,0)
					(6.25,0)
					(6.5,0)
					(6.75,0)
					(7,0)
					(7.25,0)
					(7.5,0)
					(7.75,0)
					(8,0)
					(8.25,0)
					(8.5,0)
					(8.75,0)
					(9,0)
					(9.25,0)
					(9.5,0)
					(9.75,0)
					(10,0)
					
				};	
			\end{axis}
			
			\begin{axis}[ 
				symbolic x coords={
					0,
					0.25,
					0.5,
					0.75,
					1,
					1.25,
					1.5,
					1.75,
					2,
					2.25,
					2.5,
					2.75,
					3,
					3.25,
					3.5,
					3.75,
					4,
					4.25,
					4.5,
					4.75,
					5,
					5.25,
					5.5,
					5.75,
					6,
					6.25,
					6.5,
					6.75,
					7,
					7.25,
					7.5,
					7.75,
					8,
					8.25,
					8.5,
					8.75,
					9,
					9.25,
					9.5,
					9.75,
					10,	
				},
				hide axis,
				ybar,
				bar width=5pt,
				ymin=0,
				every node near coord/.append style={rotate=90, anchor=west},
				width=\textwidth, 
				height=9cm, 
				width=16cm,
				axis x line*=bottom, axis y line*=left
				]
				\addplot [fill=yellow] coordinates {
					(0,0)
					(0.25,0)
					(0.5,0)
					(0.75,0)
					(1,0)
					(1.25,0)
					(1.5,0)
					(1.75,0)
					(2,0)
					(2.25,0)
					(2.5,0)
					(2.75,0)
					(3,0)
					(3.25,0)
					(3.5,0)
					(3.75,0)
					(4,0)
					(4.25,0)
					(4.5,0)
					(4.75,0)
					(5,0)
					(5.25,0)
					(5.5,0)
					(5.75,0)
					(6,0)
					(6.25,0)
					(6.5,0)
					(6.75,0)
					(7,0)
					(7.25,0)
					(7.5,30000)
					(7.75,0)
					(8,0)
					(8.25,0)
					(8.5,0)
					(8.75,0)
					(9,0)
					(9.25,0)
					(9.5,0)
					(9.75,0)
					(10,0)
					
				};	
			\end{axis}
			
			\begin{axis}[
				ylabel={Number of Student},
				symbolic x coords={
					0,
					0.25,
					0.5,
					0.75,
					1,
					1.25,
					1.5,
					1.75,
					2,
					2.25,
					2.5,
					2.75,
					3,
					3.25,
					3.5,
					3.75,
					4,
					4.25,
					4.5,
					4.75,
					5,
					5.25,
					5.5,
					5.75,
					6,
					6.25,
					6.5,
					6.75,
					7,
					7.25,
					7.5,
					7.75,
					8,
					8.25,
					8.5,
					8.75,
					9,
					9.25,
					9.5,
					9.75,
					10,	
				},
				xtick=data,
				x tick label style={rotate=90,anchor=east},
				ybar,
				bar width=5pt,
				ymin=0,
				ymax=22500,
				nodes near coords,   
				every node near coord/.append style={rotate=90, anchor=west},
				width=\textwidth, 
				height=9cm, 
				width=16cm,
				axis x line*=bottom, axis y line*=left
				]
				\addplot [fill=blue] coordinates {
					(0,0)
					(0.25,1)
					(0.5,7)
					(0.75,25)
					(1,117)
					(1.25,312)
					(1.5,640)
					(1.75,1329)
					(2,2427)
					(2.25,3557)
					(2.5,4868)
					(2.75,6273)
					(3,7526)
					(3.25,8596)
					(3.5,9500)
					(3.75,10708)
					(4,11611)
					(4.25,12538)
					(4.5,13917)
					(4.75,15307)
					(5,16332)
					(5.25,17434)
					(5.5,18557)
					(5.75,19414)
					(6,19656)
					(6.25,19839)
					(6.5,19393)
					(6.75,18687)
					(7,17519)
					(7.25,15238)
					(7.5,13134)
					(7.75,10326)
					(8,7821)
					(8.25,5425)
					(8.5,3408)
					(8.75,2046)
					(9,966)
					(9.25,385)
					(9.5,103)
					(9.75,17)
					(10,2)
					
				};	
				
			\end{axis}
		\end{tikzpicture}
	\end{center}
	\caption{Comparison between ChatGPT-BingChat performances and Vietnamese students in 2019.}
	\label{fig:phy_2019}
\end{figure*}
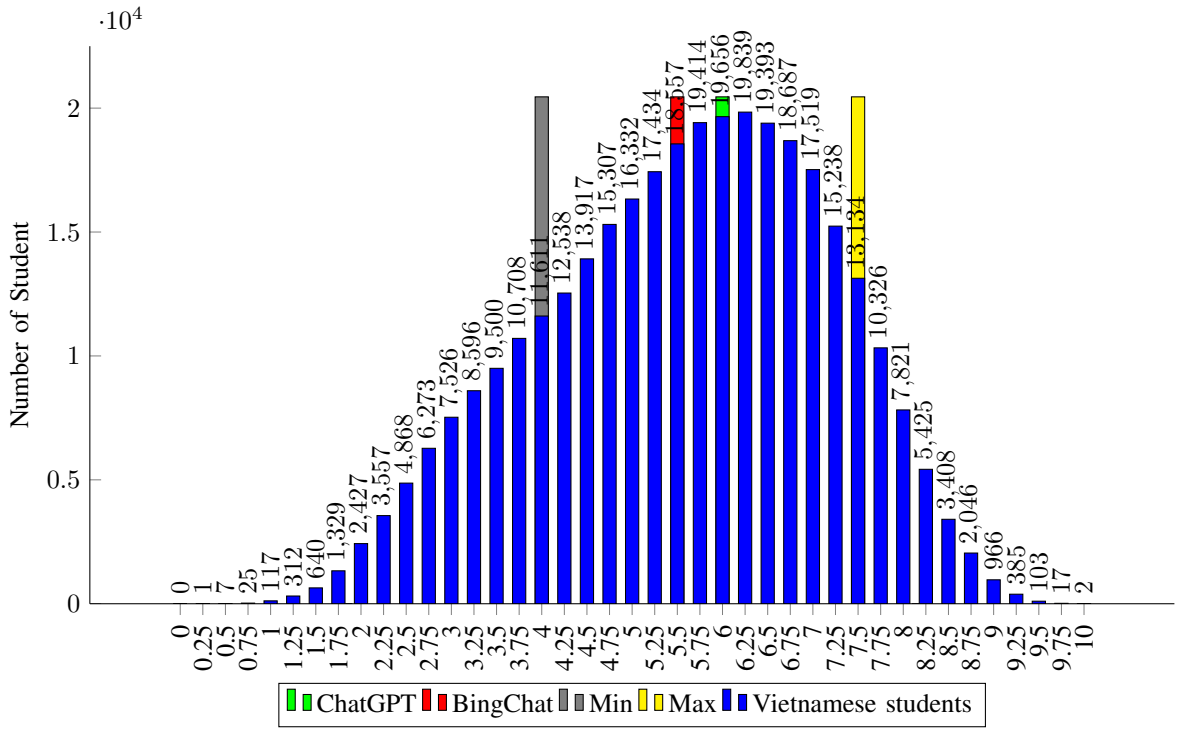

\section{Discussion}

LLMs have a great deal of potential to change education by providing individualized and interactive learning experiences. Large volumes of data may be analyzed by LLMs, which can also provide customized feedback and adapt to different learning methods. They can help teachers work less by helping to grade and appraise student work. The results of this study indicate that LLMs like ChatGPT and BingChat are less accurate than Vietnamese students and have a limited capacity to respond to high-level application problems in Physics at the high school level. This emphasizes the difficulties LLMs encounter when attempting to comprehend the intricacies of natural language, particularly in specialist fields like physics. Regardless of their location or socioeconomic situation, LLMs can give students access to excellent materials and tailored feedback despite these obstacles. 

Moreover, LLMs can also be taught to recognize and adjust to regional variations in language and culture, making them useful in a variety of settings, including Vietnam. The accuracy and reliability of LLMs in specialized fields like physics need to be improved, platforms and tools need to be created to make it easier to integrate LLMs into the classroom, and privacy and data security issues need to be resolved in order to fully realize the potential of LLMs in education. LLMs have the potential to revolutionize education, including in the field of physics in Vietnam and around the world, but their success depends on resolving the issues this study has brought to light and putting in place the policies and infrastructure necessary for their successful integration into the educational system.

\
\section{Conclusion}

Our study's objective was to evaluate how well ChatGPT and BingChat, two SOTA-LLMs, performed when answering high school physics questions in exams given in Vietnamese between 2019 and 2023. The findings showed that both LLMs had trouble answering complex application questions. BingChat showed better accuracy, while ChatGPT was more reliable in its responses. Our investigation also compared the LLMs' performance to the test results of Vietnamese students, and the results showed that ChatGPT and BingChat performed worse than Vietnamese students. This demonstrates the limitations of LLMs as a substitute for human intellect in the teaching of physics. However, LLMs can still help students and teachers with individualized instruction and provide immediate feedback. Additionally, they can produce practice and test materials. However, more domain-specific knowledge must be incorporated into LLMs in order to improve their ability to reason and apply knowledge. Future studies should concentrate on enhancing LLMs' capability to answer complex questions and assessing how well they contribute to improved student learning outcomes.

\bibliographystyle{IEEEtran}
\bibliography{paper}

\end{document}